\documentstyle[12pt,epsfig]{article} 
\def\baselinestretch{1.3}
\newcommand{\ba}{\begin{array}}
\newcommand{\ea}{\end{array}}
\newcommand{\bd}{\begin{displaymath}}
\newcommand{\ed}{\end{displaymath}}
\newcommand{\be}{\begin{equation}}
\newcommand{\ee}{\end{equation}}
\newcommand{\bea}{\begin{eqnarray}}
\newcommand{\eea}{\end{eqnarray}}
\newcommand{\sla}[1]{/\!\!\!#1}

\def\a{\alpha}

\def\q2 {q^2}

\def\r {\rightarrow}

\def\miss {\hspace{-0.5cm}\slash~~}

\def\rslep {\tilde{e_R}}
\def\rsnu {\tilde{\nu}_R}
\def\snu {\tilde{\nu}}
\def\lslep {\tilde{e_L}}
\def\stau {\tilde{\tau}}
\def\mer {m_{\rslep}}
\def\mmr {m_{\tilde{\mu}_R}}
\def\mml {m_{\tilde{\mu}_L}}
\def\mel {m_{\lslep}}

\def\bt{\begin{table}}
\def\et{\end{table}}

\catcode`@=11 
\def \gsim{\mathrel{\mathpalette\@versim>}}
\def \lsim{\mathrel{\mathpalette\@versim<}}
\def \@versim#1#2{\lower0.4ex\vbox{\baselineskip\z@skip\lineskip\z@skip
     \lineskiplimit\z@\ialign{$\m@th#1\hfil##\hfil$%
     \crcr#2\crcr\sim\crcr}}}
\catcode`@=12 

\begin{document}

\begin{flushright}
{RECAPP-HRI-2009-019}
\end{flushright}

\begin{center}

{\large\bf Chargino reconstruction in supersymmetry with long-lived staus}\\[15mm]
Sanjoy Biswas\footnote{E-mail: sbiswas@mri.ernet.in}
and Biswarup Mukhopadhyaya\footnote{E-mail: biswarup@mri.ernet.in}\\
{\em Regional Centre for Accelerator-based Particle Physics \\
     Harish-Chandra Research Institute\\
Chhatnag Road, Jhunsi, Allahabad - 211 019, India}
\\[20mm] 
\end{center}

\begin{abstract} 
We consider a supersymmetric (SUSY) scenario including right-handed
neutrinos, one of whose scalar superpartners  is the lightest
SUSY particle (LSP). The distinguishing feature in the collider signal of SUSY
 in such a case is not missing energy but a pair of charged tracks
corresponding to the next-to-lightest SUSY particle, when it is,
as in the case considered, a stau.
Following up on our recent work on neutralino reconstruction in such cases,
we explore the possibility of reconstructing charginos, too, through a
study of transverse mass distributions in specified final states. The
various steps of isolating the transverse momenta of neutrinos relevant
for this are outlined, and regions of the parameter space where
our procedure works are identified.
\end{abstract}

\vskip 1 true cm

\newpage
\setcounter{footnote}{0}

\def\baselinestretch{1.5}
\section{Introduction}

If supersymmetry (SUSY) \cite{Book,Sally}, broken at the TeV scale, has to
validate itself as the next step in physics beyond the standard model
(SM), then it is likely to be discovered at the Large Hadron Collider
(LHC), with the superpartners of the SM particles being identified.
It is therefore of great importance to make a thorough inventory of
collider signals answering to various SUSY scenarios.  These can serve
not only to unveil the general character of the scenario but also to
yield a wealth of information as possible about the specific
properties of the new particles.

One attractive feature of supersymmetric theories, in the $R$-parity
(defined as $R=(-)^{3B+L+2S}$) conserving form, is that the
superparticles are always produced in pairs and each interaction vertex
must involve an even number of superparticles (having R = -1). Hence
the lightest SUSY particle (LSP) is stable and all SUSY cascades at
the collider experiments should end up with the pair production of the
LSP. As a bonus, it provides a viable candidate for dark matter (DM)
if the LSP is electrically neutral and only weakly interacting. Since
the LSP, due to such a character, escapes the detector without being
detected, a prototype signature of $R$-conserving SUSY is energetic
jets and/or leptons associated with large missing transverse energy
($\sla E_T$).

However, there can be several SUSY models \cite{Hamaguchi:2006vu} 
which include a quasi-stable
next-to-lightest supersymmetric particle (NLSP). This can happen in
cases where the NLSP is nearly degenerate in mass with the LSP, or if
its coupling to the LSP is too small. Hence the decay width of the
NLSP into the LSP is highly suppressed. Consequently the NLSP becomes
stable on the detector scale, its lifetime being long enough to
escape the detector without decaying inside it. Thus the NLSP behaves
like a stable particle within the detector.  The resulting collider
signals change drastically, especially if the NLSP is a charged
particle. The quintessential SUSY signals then is not $\sla E_T$ but
two hard charged tracks of massive stable particles which appear as
far as up to the muon chamber. This opens up a whole set of new
possibilities for collider studies, including reconstruction of the
sparticle masses, something that is relatively more difficult in the
presence of $\sla E_T$.

The scenario we have considered here, as an illustration of such a
quasi-stable NLSP, is the minimal supersymmetric standard model (MSSM)
augmented with a right-chiral Dirac-type neutrino superfields for each
generation. This is consistent with the existing evidence \cite{nudata} of neutrino
masses and mixing, although no explanation for the smallness of
neutrino masses is offered. It is possible in such a case to have an
LSP is dominated by this right-chiral sneutrino
state ($\snu_{R}$) together with a charged particle as the NLSP. We
specifically consider a situation with a stau ($\tilde{\tau}$) NLSP
\footnote{It should be
remembered that the above possibility is not unique. One may as well have 
a spectrum in which a third family squark is the NLSP \cite{Chou:1999zb,Choudhury:2008gb}.}. 
Such a scenario can easily be motivated \cite{sakurai} by assuming that the
MSSM is embedded in a high scale framework of SUSY breaking. As we
shall see, this can happen in minimal supergravity (mSUGRA) \cite{mSUGRA} where the
masses evolve from ``universal'' scalar ($m_0$) and gaugino ($M_{1/2}$)  
mass parameters at a high scale. The only extension here is
the right-chiral neutrino superfield (in fact, three of them) 
whose scalar component derives
its soft mass from the same $m_0$. The existence of such a quasistable 
charged slepton can be well in agreement with the observed
abundance of  light elements as predicted by the big-bang
nucleosynthesis (BBN) \cite{BBN2007}, provided, its mass is below a TeV.
Since the right-chiral sneutrino has no gauge couplling but only 
interactions proportional to the neutrino 
Yukawa coupling, the strength of which is too feeble to be
seen in dark matter search experiments, such an LSP is consistent with
all direct searches carried out so far. Moreover, it has been shown that such a
spectrum is consistent with all  low energy constraints \cite{Amsler:2008zzb}, 
and the contribution to the relic density of the universe 
can be compatible with the limits set by
the Wilkinson Microwave Anisotropy Probe (WMAP) \cite{wmap} with
appropriate values of the relevant parameters \cite{Asaka:2005cn}.

In this work, we have concentrated on the mass reconstruction of the
lightest chargino in a $\stau$-NLSP, $\snu_1$-LSP scenario. This is a
follow-up of our earlier work \cite{Biswas:2009zp} on neutrlino
reconstruction under similar circumstances. We have shown that
it is possible to
determine the mass of the lightest chargino, produced in the cascade
decay of squarks or gluinos, from the sharp drop noticed in the
transverse mass \cite{TNVM:1983a} distribution of the chargino decay
products. More precisely, we show a way of disentangling the transverse
mass of the system consisting of a $\stau$-track and the associated
neutrino from chargino decay. We suggest a method for extracting the
transverse momentum ($p_T$) of the neutrino. Though a sizable statistics is required
for this purpose, and one may have to wait for considerable
accumulated luminosity after the discovery of the LHC, still this is
a rather spectacular prospect.  We have successfully applied the
criteria, developed in our earlier work, for separating the signal
from standard model backgrounds. Ways of suppressing SUSY processes
that are likely to contaminate the transverse mass distributions are
also suggested.

It should be mentioned
that neither the signal we have studied nor the prescribed reconstruction
technique is limited just to scenarios with right-sneutrino LSP.
It can be applied successfully to all cases \cite{gravitinoLSP,gmsbnlsp,coanni} where the
NLSP is a charged scalar with quasi-stable character, provided that
it decays outside the detector, leaving behind a charged track in the
muon chamber.

The paper is organized as follows: in Sec. 2, we motivate the
scenario under investigation and present a brief review of the mass
reconstruction for neutralinos as done in the earlier paper and used
in this work. There we also summarise our choice of benchmark points. The
signal under study and the reconstruction strategy for determining the
chargino mass as well as the possible sources of background, both from
the SM and within the model itself, and their possible discrimination,
are discussed in Sec. 3. We summarise and conclude in Sec. 4.


\section{The overall scenario,  mass reconstruction and representative  benchmark points} 

\subsection{Scenarios with $\snu_{R}$ LSP and $\stau$ NLSP}

As has been already stated, the most simple-minded extension of the
MSSM \cite{S.P.Martin1}, accommodating neutrino masses, is the addition of one right-handed
neutrino superfield per family. In this situation the neutrinos have
Dirac masses induced by very small Yukawa couplings.
The superpotential of such an  extended MSSM becomes
(suppressing family indices),

\be W_{MSSM} = y_l
\hat{L}\hat{H_d}\hat{E^c} + y_d \hat{Q}\hat{H_d}\hat{D^c}+y_u \hat{Q}
\hat{H_u} \hat{U^c}+\mu\hat{H_d}\hat{H_u}+y_\nu \hat{H}_u \hat{L}\hat{\nu}^c_R 
\ee
\noindent
where $\hat{H_d}$ and $\hat{H_u}$, respectively, are the Higgs
superfields that give masses to the $T_{3}=-1/2$ and
$T_{3}=+1/2$ fermions, and $y's$ are the strengths of Yukawa
interaction. $\hat{L}$ and $\hat{Q}$ are the left-handed lepton and
quark superfields respectively, whereas $\hat{E^c}$, $\hat{D^c}$ and
$\hat{U^c}$, in that order, are the right handed gauge singlet charged
lepton, down-type and up-type quark superfields. $\mu$ is the Higgsino
mass parameter.

It is a common practice to attempt reduction of free parameters in the
theory, by assuming
a high-scale framework of SUSY breaking. The most commonly
adopted scheme is based on N = 1 minimal supergravity (mSUGRA). 
There SUSY breaking in the hidden sector at high scale is
manifested in universal soft masses for scalars ($m_0$) and gauginos
($M_{1/2}$), together with the trilinear ($A$) and bilinear ($B$) SUSY
breaking parameters in the scalar sector. The bilinear parameter is
determined by the electroweak symmetry breaking (EWSB) conditions.
All the scalar and gaugino masses at
low energy obtained by renormalization group evolution (RGE) of the
universal mass parameters $m_{0}$ and $M_{1/2}$ from high-scale
values \cite{S.P.Martin2}. 
Thus one generates all the squark, slepton, and gaugino masses as well as all the
mass parameters in the Higgs sector. The Higgsino mass
parameter $\mu$ (up to a sign), too is determined from EWSB
conditions. All one has to do in this scheme is to specify  the high 
scale ($m_0, M_{1/2}, A_0$, together with 
$sign(\mu)~{\rm and}~\tan\beta ~=~ \langle H_u \rangle/  \langle H_d \rangle$)
where, $\tan\beta$ is the ratio of the vacuum expectation values of the
two Higgs doublets that give masses to the up-and down-type quarks
respectively.

The neutrino masses are typically given by,
\bea
m_\nu = y_\nu \left<H_u^0\right> = y_\nu v~\sin\beta
\eea
\noindent
The small Dirac masses of the neutrinos imply that the
neutrino Yukawa couplings ($y_{\nu}$) are quite small ($\lsim 10^{-13}$).

With the inclusion of the right-chiral neutrino superfields as a
minimal extension, it makes sense to assume
that the masses of their scalar components, too, 
originate in the same parameter $m_{0}$. The
evolution of all other parameters practically remain the same in this
scenario as in the MSSM, while the right-chiral sneutrino mass
parameter evolves at the one-loop level \cite{arkani} as
\bea
\frac{dM^2_{\rsnu}}{dt} = \frac{2}{16\pi^2}y^2_\nu~A^2_\nu ~~.
\eea
\noindent
where $A_{\nu}$ is obtained by the running of the trilinear soft SUSY
breaking term $A$ and is responsible for left-right mixing in the
sneutrino mass matrix.

It follows from above that the value of $M_{\rsnu}$ remains practically 
frozen at $m_{0}$, thanks to the extremely small Yukawa couplings,
whereas the other sfermion masses are enhanced at the electroweak
scale. Thus, for a wide range of values of the gaugino masses, one 
naturally ends up with a sneutrino LSP ($\snu_{1}$), dominated by the
right-chiral state. This is because the mixing angle is controlled by
the neutrino Yukawa couplings:
\bea
\tilde{\nu}_1 = - \tilde{\nu}_L \sin\theta + \tilde{\nu}_R \cos\theta
\eea
\noindent
where the mixing angle $\theta$ is given by,
\bea
\tan 2\theta = \frac{2 y_\nu v\sin\beta |\cot\beta\mu -
A_\nu|}{m^2_{\tilde{\nu}_L}-m^2_{\rsnu}}
\eea

Of the three charged sleptons, the amount of left-right mixing is
always the largest in the third, and hence the lighter stau
($\stau_1$) often turns out to be the NLSP in such a scenario. There
are regions in the parameter space the three lighter sneutrino states
corresponding to the three flavours act virtually as co-LSP's. It is,
however, sufficient for illustrating our points to consider the
lighter sneutrino mass eigenstate of the third family, as long as the
state ($\stau_1$) is the lightest among the charged sleptons. Thus the
addition of a right-handed sneutrino superfield, for each family,
which is perhaps the most minimal input to explain neutrino masses and
mixing, can eminently turn a mSUGRA theory into one with a stau NLSP
and a sneutrino LSP. It is should be emphasized that the physical LSP
state can have (a) Yukawa couplings proportional to the neutrino mass,
and (b) gauge coupling with the small left-chiral admixture in it,
driven by left-right mixing which is again proportional to
$y_{\nu}$. Thus the decay of any particle (particularly the NLSP) into
the LSP will always be a very slow process, not taking place within
the detector. Under such circumstances, the quintessential SUSY signal
is not $\sla{E_T}$ anymore but a pair of charged tracks left by the
quasi-stable NLSP.

\subsection{Neutralino reconstructed}

In an earlier study \cite{Biswas:2009zp}, we suggested a reconstruction technique for 
at least one of the two lightest neutralinos, 
in the $\snu_{R}$ LSP and $\stau$ NLSP scenario. The
signal studied there was:  $2\tau_j+2\stau (charged-track)+E_{T}\miss+X$.
Here $\tau_j$ denotes a jet out of one-prong decays of the tau,
and all accompanying hard jets arising from cascades are included in X.
The kinematic cuts imposed by us, such as 
$p^{track}_T>$100 GeV and $\sum\vec|p_T|>$1 TeV, ($p_T$ = transverse momentum,
and $\sum\vec|p_T|$ is the  scalar sum of all visible tarnsverse momenta)
reduced the backgrounds considerably.

Since the signal we investigated involves two taus in the final state, and hadronic
decays of tau was considered, tau-jet identification and tau reconstruction 
were two important components of the procedure. 
For this a method suggested in \cite{Rainwater:1998kj} was used, which involved
solving the following equation event-by-event:
\be
\vec\sla{p_T} = ({1\over x_{\tau_{h1}}} - 1) \; \vec p_{h1} +
({1\over x_{\tau_{h2}}} - 1) \; \vec p_{h2} \; 
\ee
\noindent
where $x_{\tau_{hi}}$ (i = 1,2) is the fraction of the tau energy
carried by each product jet collinear with the parent tau, when it is
boosted. $\vec\sla{p_T}$ is the vector sum of the transverse
components of the 3-momenta of the two product neutrinos produced in
hadronic decay of each tau. Clearly this method is applicable when
there is no other invisible particle in the final state. This was
ensured in the best possible manner by vetoing any isolated lepton
in the final state, thus getting rid of additional neutrinos from W-decays.

The reconstruction of $\vec\sla{p_T}$ is undoubtedly  very crucial here.
$\vec\sla{p_T}$ is
reconstructed as the negative of the total visible $\vec{p_T}$ which
receives the contributions from isolated
leptons/ sleptons, jets and unclustered components. The last among these
includes all particles (electron/photon/ muon/stau) with
$0.5<E_T< 10$GeV and $|\eta|< 5$ (for muon or muon-like tracks,
$|\eta|< 2.5$), or hadrons with $0.5< E_T< 20$GeV and $|\eta|<5$,
which do not contribute to a jet and constitute `hits' in the detector \cite{TDR}. 
In order to simulate the
finite resolution of detectors, the energies/transverse momenta of
all particles were smeared following prescriptions detailed in \cite{Biswas:2009zp}.

Once the tau 4-momenta are obtained, they are combined with the
staus to find the stau-tau invariant mass distribution. This requires
the knowledge of stau mass\footnote{From the muon chamber only the
three-momenta of the charged track can be obtained.} as well as the
choice of the correct pair. The correct pairs are obtained by using a
seed mass for $\stau$ which was taken to be 100 GeV, satisfying the criterion
$|M^{pair1}_{\stau\tau}-M^{pair2}_{\stau\tau}|<50 ~GeV$. The actual
stau mass is then extracted by demanding that the invariant masses of
the two $\stau \tau$ pairs were equal, which yields an equation
involving one unknown quantity, namely, $m_{\stau}$:
\be
\sqrt{m^2_{\stau}+|\vec{p_{\stau_1}}|^2}.E_{\tau_1}-\sqrt{m^2_{\stau}+
|\vec{p_{\stau_2}}|^2}.E_{\tau_2}
=\vec{p_{\stau_1}}.\vec{p_{\tau_1}}-\vec{p_{\stau_2}}.\vec{p_{\tau_2}}
\ee

Having thus extracted $m_{\stau}$ on an event-by-event basis in the
event generator, it was demonstrated that the distribution of this
mass value has a peak at the actual $m_{\stau}$. We have used this peak
value in reconstructing the neutralino from the invariant mass
distribution of the $\stau \tau$ pair. In some regions of the parameter space,
it is possible to thus reconstruct only $\chi^0_1$, as the production
rate of $\chi^0_2$ in cascade decay of $\tilde q$ (or $\tilde g$) as
well as the decay branching ratio of $\chi^0_{2} \r \stau\tau$ is
small. In some other regions, we have been able to reconstruct
both of them. There are still other regions only the $\chi^0_2$ peak 
shows up. This is because of the small
mass splitting between $\chi^0_1$ and $\stau$, which softens the tau
(jet) arising from its decay, preventing it from passing the requisite
hardness cut.

\subsection{The choice of benchmark points}

The choice of benchmark points for this study is the same as in the
case of neutralino reconstruction \cite{Biswas:2009zp}. The mSUGRA parameter space is
utilised for this purpose. A $\stau$ NLSP and a $\tilde \nu_1$ LSP occur in
those regions in which one would have had a $\stau$ LSP in the absence
of right-chiral neutrino superfields. We focus on both the regions
where (a) $m_{{\stau}_1} > m_{{\tilde \nu}_1} + m_W$, and (b) the
above inequality is not satisfied. In the first case, the dominant
decay mode is the two-body decay of the NLSP, ${\stau}_1 \to
\tilde{\nu}_1 W$, and, in the second, the decay takes place via a
virtual $W$. Decay into a charged Higgs is a subdominant channel for
the lighter stau.  The decay takes place outside the detector in all
cases. At different benchmark points, however, the mass splittings
between the $\stau_1$ and neutralinos/charginos are different. This in
turn affects the kinematic characteristics of the final states under
consideration.

We have used the spectrum generator {\bf \begin{footnotesize}ISAJET 7.78\end{footnotesize}}
\cite{isajet} for our study. In Table-1 we list the six benchmark
points used, both in terms of high-scale parameters and low-energy
spectra. The justification of their choice and their representative
character have been explained in reference \cite{Biswas:2009zp}.

\begin{table}[htb]
\begin{tabular}{||c||c|c|c|c|c|c||}
\hline
\hline
       & {\bf BP-1}&{\bf BP-2}&{\bf BP-3}&{\bf BP-4}&{\bf BP-5}&{\bf BP-6} \\
\hline
 mSUGRA     &$m_0=100$&$m_0=100$&$m_0=100$&$m_0=100$&$m_0=100$
            &$m_0=100$\\
 input      &$m_{1/2}=600$&$m_{1/2}=500$&$m_{1/2}=400$&$m_{1/2}=350$
            &$m_{1/2}=325$&$m_{1/2}=325$\\
            &$\tan\beta=30$ &$\tan\beta=30$&$\tan\beta=30$&$\tan\beta=30$&$\tan\beta=30$
&$\tan\beta=25$\\
\hline
$\mel,\mml$   &418&355&292&262&247&247\\
$\mer,\mmr$   &246&214&183&169&162&162\\
$m_{\snu_{eL}},m_{\snu_{\mu L}}$&408&343&279&247&232&232\\
$m_{\snu_{\tau L}}$&395&333&270&239&224&226\\
$m_{\snu_{iR}}$&100&100&100&100&100&100\\
$m_{\stau_1}$&189&158&127&112&106&124\\
$m_{\stau_2}$&419&359&301&273&259&255\\
\hline
$m_{\chi^0_1}$&248&204&161&140&129&129\\
$m_{\chi^0_2}$&469&386&303&261&241&240\\
$m_{\chi^{\pm}_1}$&470&387&303&262&241&241\\
$m_{\tilde{g}}$&1362&1151&937&829&774&774\\
$m_{\tilde{t}_1}$&969&816&772&582&634&543\\
$m_{\tilde{t}_2}$&1179&1008&818&750&683&709\\
$m_{h^0}$ &115&114&112&111&111&111\\
\hline
\hline
\end{tabular}\\
\caption {\small \it Proposed benchmark points (BP) for the 
study of the stau-NLSP scenario
in SUGRA with right-sneutrino LSP. The values of $m_0$ and
$M_{1/2}$ are given in GeV. We have also set $A_0=100~GeV$ and
$sgn(\mu)=+$ for benchmark points under study.}
\label{tab:1}       
\end{table}

It may be noted that the region of the mSUGRA parameter space where we
have worked is consistent with all the experimental bounds \cite{constraints}, including
both collider and low-energy constraints (such as the LEP and Tevatron
constraints on the masses of Higgs, gluinos, charginos etc.,  as
also those from $b \rightarrow s\gamma$, correction to the
$\rho$-parameter, ($g_\mu - 2$) and so on). In the next section we 
describe the procedure for the reconstruction of $\chi^{\pm}_1$,
for these benchmark points.

\section{Reconstruction of the lighter chargino}

The final state of use for the reconstruction of the lighter chargino is 

\begin{itemize}
\item $\tau_j+2\stau {\rm (opposite-sign~charged~tracks)}+E_{T}\miss+X$
\end{itemize}
where, $\tau_j$ represents a jet which has been identified as a tau
jet, the missing transverse energy is denoted by $E_{T}\miss$ and all
other jets coming from cascade decays are included in X \footnote{In
order to avoid the combinatorial backgrounds, we have considered
events with two opposite sign $\stau$'s only.}.

Simulation for the LHC has been done for the signal as well as
backgrounds using {\bf \begin{footnotesize}PYTHIA\end{footnotesize}} 
(v6.4.16) \cite{PYTHIA}.  The $pp$ events has
been studied with a centre-of-mass energy ($E_{c.m.}$)=14 TeV at an
integrated luminosity of 300 $fb^{-1}$. The numerical values of the 
electromagnetic and the strong coupling constant have been set at 
$\a^{-1}_{em}(M_Z)=127.9$ and  $\a_{s}(M_Z)=0.118$ respectively 
\cite{Amsler:2008zzb}. The hard scattering process
has been folded with CTEQ5L parton distribution function \cite{Lai:1999wy}.  We
have set the factorization and renormalization scales at the average
mass of the particles produced in the parton level hard scattering
process. In order to make our estimate
conservative, the signal rates have not been multiplied by any
K-factor \cite{ksusy}. The effects of initial and final state radiation as well as
the finite detector resolution of the energies/momenta of the final state
particles have been taken into account.

\subsection{Chargino reconstruction from transverse mass distribution}

Now we are all set to describe the main principle adopted by us for
chargino ($\chi^{\pm}_1$) reconstruction.  For this purpose, we have
looked for the processes in which $\chi^{\pm}_1-\chi^0_1/\chi^0_2$ is
being produced in association with hard jets in cascade decays of
squarks and gluinos. The $\chi^{\pm}_1$ subsequently decays into a
$\stau$-$\nu_{\tau}$ pair, while the $\chi^0_1$ (or $\chi^0_2$) decays
into a $\stau$-$\tau$ pair. Since the decay of $\chi^{\pm}_1$ involves
an invisible particle ($\nu_{\tau}$), for which it is not possible to
know all the four components of momenta, a transverse mass
distribution, rather than invariant mass distribution, of
$\stau$-$\nu_{\tau}$ pair will give us mass information of
$\chi^{\pm}_1$. In spite of the recent progress in measuring the
masses of particles in semi-invisible decay mode (for example the
$m_{T_2}$ variable introduced by \cite{Lester:1999tx} and its further implications
\cite{Gripaios:2007is}), we have focused on transverse mass variable ($m_{T}$)
because of the fact that the only invisible particle present in the
final state is the neutrino, which is massless.

The procedure, however, still remains problematic, because the $\tau$
on the other side (arising from neutralino decay) also produces a
neutrino in the final state, which contributes to $E_{T}\miss$. In
order to correctly reconstruct the transverse mass of the $\stau -
\nu_{\tau}$ pair from chargino decay, the contribution to $E_{T}\miss$
from the aforementioned neutrino must be subtracted.

Keeping this in mind, we have prescribed a method for reconstruction
of the transverse component of the neutrino 4-momenta ($\vec
P^T_{\nu_1}$) produced from the decay of $\chi^{\pm}_1$, in
association with $\stau$. To describe it in short:

We label the transverse momentum of the neutrino coming from chargino
decay by $\vec P^T_{\nu_1}$. Attention is focused on cases where the
tau, produced from a neutralino, decays hadronically and the
$\tau$-jet, out of a one-prong decay of tau, is identified following
the prescription of \cite{Rainwater:1998kj}. We have assumed a true tau-jet
identification efficiency to be 50$\%$, while a non-tau jet rejection
factor of 100 has been used \cite{Coadou:tauidfic,CMS1,Asai:2004ws} 
(The results for higher
identification efficiency are also shown in Sec. 4.). We have also
assumed that there is no invisible particle other than the two
neutrinos mentioned above.  We have attempted to ensure this by
vetoing any event with isolated charged leptons. This only leaves out
neutrinos from $Z$-decay and $W$-decays into a $\tau \nu_{\tau}$
pair. The contamination of our signal from these are found to be
rather modest.

The
transverse momenta of the neutrino ($\vec P^T_{\nu_2}$), out of a tau
decay, is first reconstructed in the collinear approximation, where 
the product neutrino and the jet are both assumed to move collinearly with the
parent tau. In this approximation, one can write
\be
P_{\tau_j}=xP_{\tau}
\ee
Following the decay $\chi^0_1$ (or, $\chi^0_2$)$\r
\stau^{\pm}\tau^{\mp}$ we have then combined the identified tau-jet with the
oppositely charged stau (track), thus forming the invariant mass
\be
m^2_{\chi^0_i}=(P_{\stau}+P_{\tau})^2=(P_{\stau}+P_{\tau_j}/x)^2   ~~~~ (i=1,2)
\ee

This pairing requires the charge information of the tau-induced jet. We
have assumed that, for a true tau-jet, the charge identification
efficiency is 100\%, while to a non-tau jet we have randomly
assigned positive and negative charge, each with 50\% weight.
One can solve this equation for $x$ (neglecting the
tau-jet invariant mass\footnote{This approximation is not valid for
  small $x$, say $x<0.1$. However, the jet out of a tau decay almost
  always carries a larger fraction of $\tau$-energy, thus justifying
  the approximation.}), to obtain

\be
x\approx\frac{2P_{\stau}.P_{\tau_j}}{m^0_{\chi^0_i}-m^0_{\stau}}  
\ee

This requires the information of $m_{\chi^0_1}$ (or, $m_{\chi^0_2}$)
and $m_{\stau}$ as well, which we have used from our earlier work for
the respective benchmark points. Once $x$ is known we have,
\be
\vec P^T_{\nu_2}=\vec P^T_{\tau}-\vec P^T_{\tau_j}=\frac{1-x}{x}.\vec P^T_{\tau_j}
\ee

Hence, the transverse component of the neutrino, out of
$\chi^{\pm}_1$-decay can be extracted from the knowledge of
$\vec \sla E_T$ of that particular event\footnote{For details 
on the reconstruction of  $\vec\sla E_T$ see
\cite{Biswas:2009zp}.}. This is given by,
\be
\vec P^T_{\nu_1}=\sla \vec E_T-\vec P^T_{\nu_2}
\ee
Finally, from the end point of the transverse mass distribution of the
$\stau$-$\nu_{\tau}$ pair the value of $m_{\chi^{\pm}_{1}}$ can be
obtained. However, one should keep in mind that, both $\chi^0_1$ or
$\chi^0_2$ can decay into a $\stau \tau$ pair. Therefore, it
is necessary to specify some criterion to separate whether a given
$\stau \tau$ pair has originated from a $\chi^0_1$ or
$\chi^0_2$ which we have discussed in the next subsection.

\subsection{Distinguishing between decay products of  $\chi^0_1$ and  $\chi^0_2$}

In order to identify the origin of a given opposite sign $\stau$-$\tau$ pair,
the first information that is to be extracted from data is
which benchmark region one is in. We have assumed of gaugino mass
universality for this process, for the sake of simplicity.

If one looks at the effective mass (defined by $M_{eff}=\sum |\vec
p_T|+\sla \vec E_T$) distribution of the final state, then the peak of
the distribution gives one an idea of the masses of the strongly interacting
superparticles which are the dominant products of the initial hard scattering
process. This is seen from
Figure 1. Once the order of magnitude of the gluino mass is inferred from
this distribution, one can use 
the universality of gaugino masses, which, in turn, 
indicates where $m_{\chi^0_1}$ and $m_{\chi^0_2}$, masses of the two
lightest neutralinos, are expected to lie.

\begin{figure}[htbp]
\begin{center}
\centerline{\epsfig{file=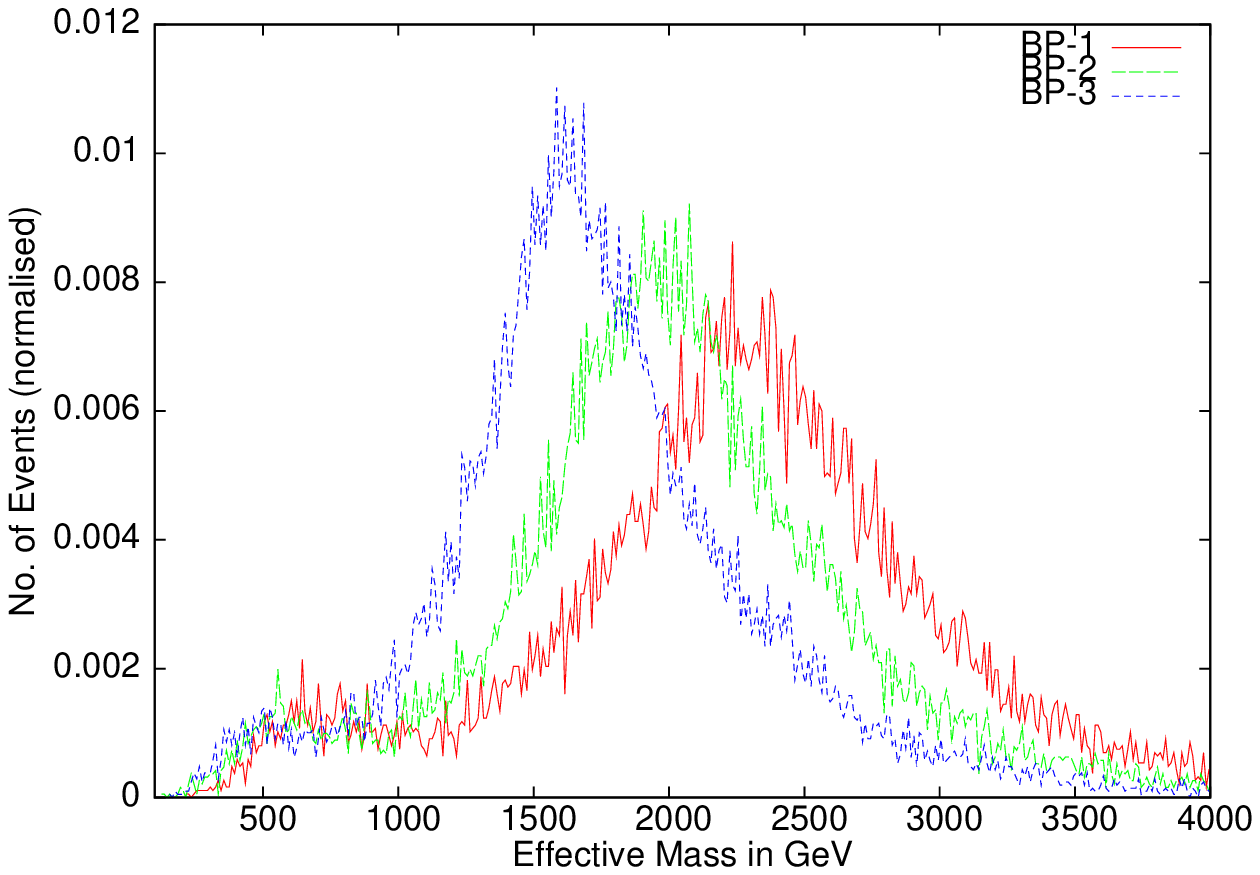,width=7.0cm,height=6.0cm,angle=-0}
\hskip 20pt \epsfig{file=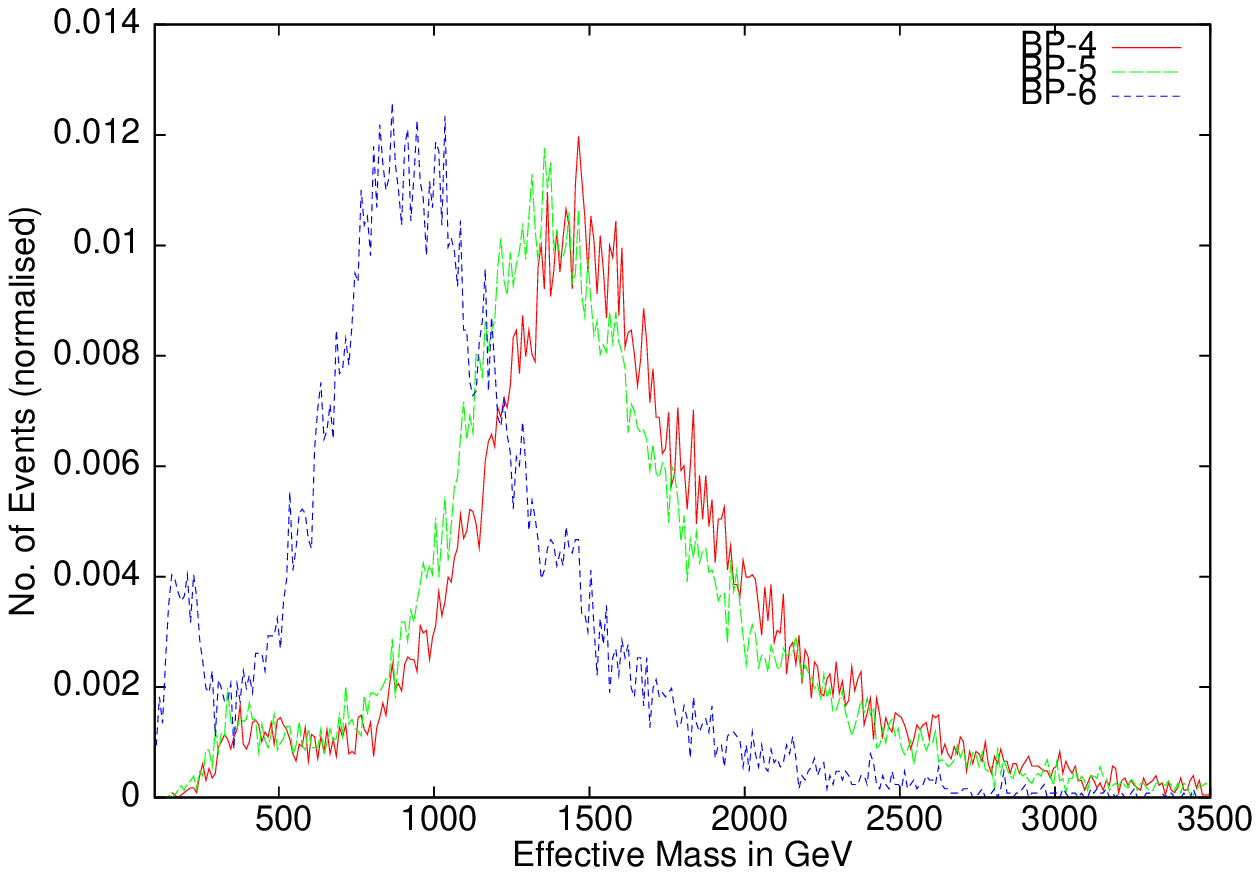,width=7.0cm,height=6.0cm,angle=-0}}
\vskip 15pt
\caption{\small \it {$M_{eff}$ distribution (normalized to unity) of the final state 
under consideration, for all benchmark points.}} 
\end{center}
\end{figure}


\begin{figure}[tbhp]
\begin{center}
\centerline{\epsfig{file=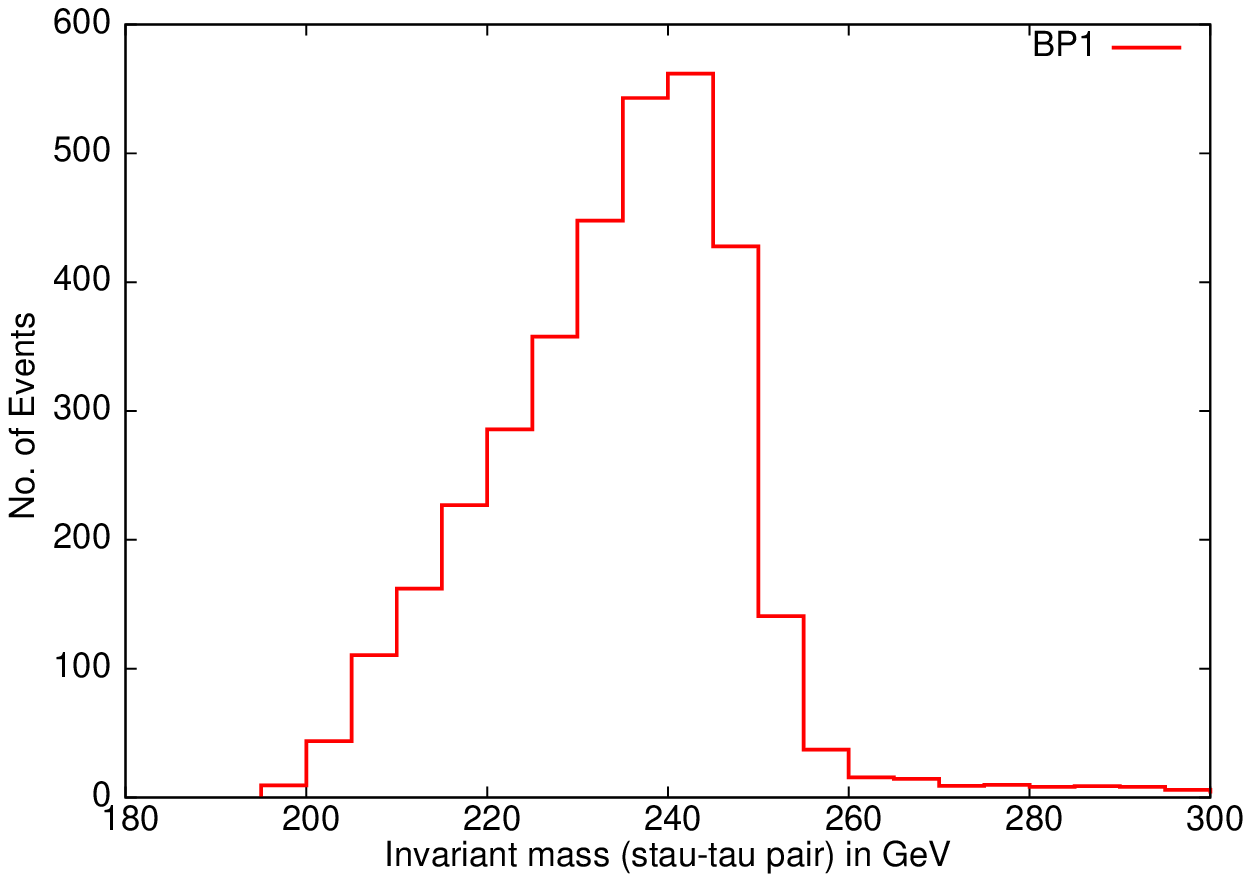,width=7.0cm,height=6.0cm,angle=-0}
\hskip 20pt \epsfig{file=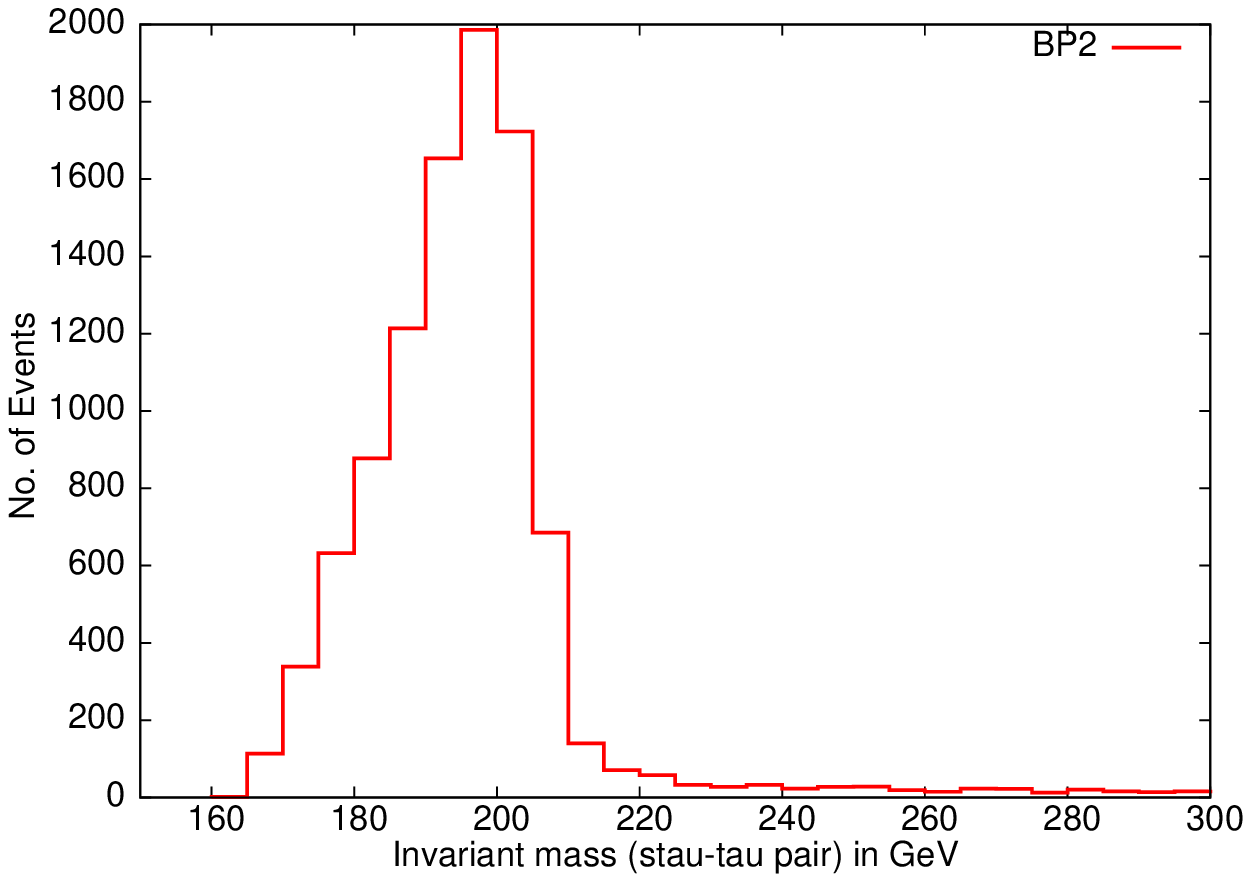,width=7.0cm,height=6.0cm,angle=-0}}
\vskip 10pt
\centerline{\epsfig{file=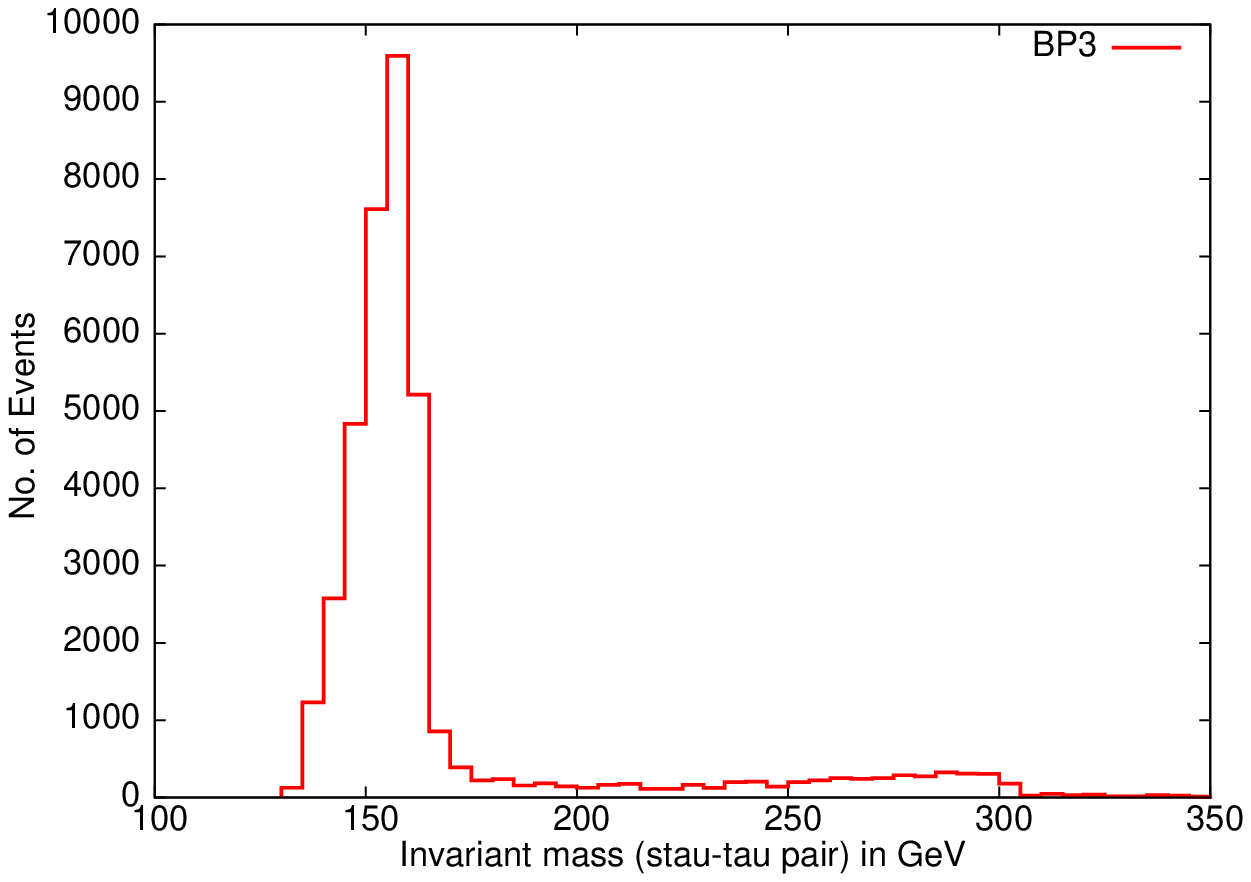,width=7.0cm,height=6.0cm,angle=-0}
\hskip 20pt \epsfig{file=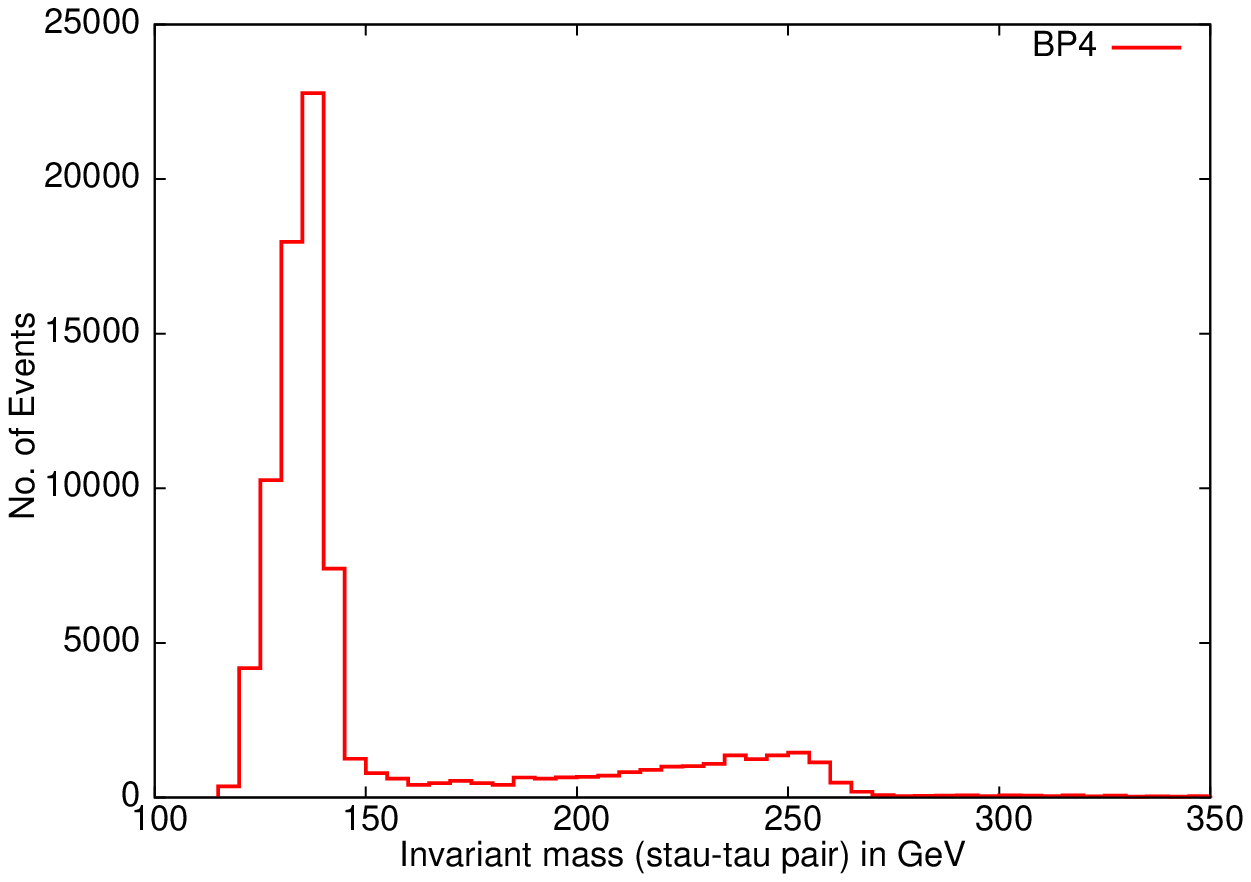,width=7.0cm,height=6.0cm,angle=-0}}
\vskip 10pt
\centerline{\epsfig{file=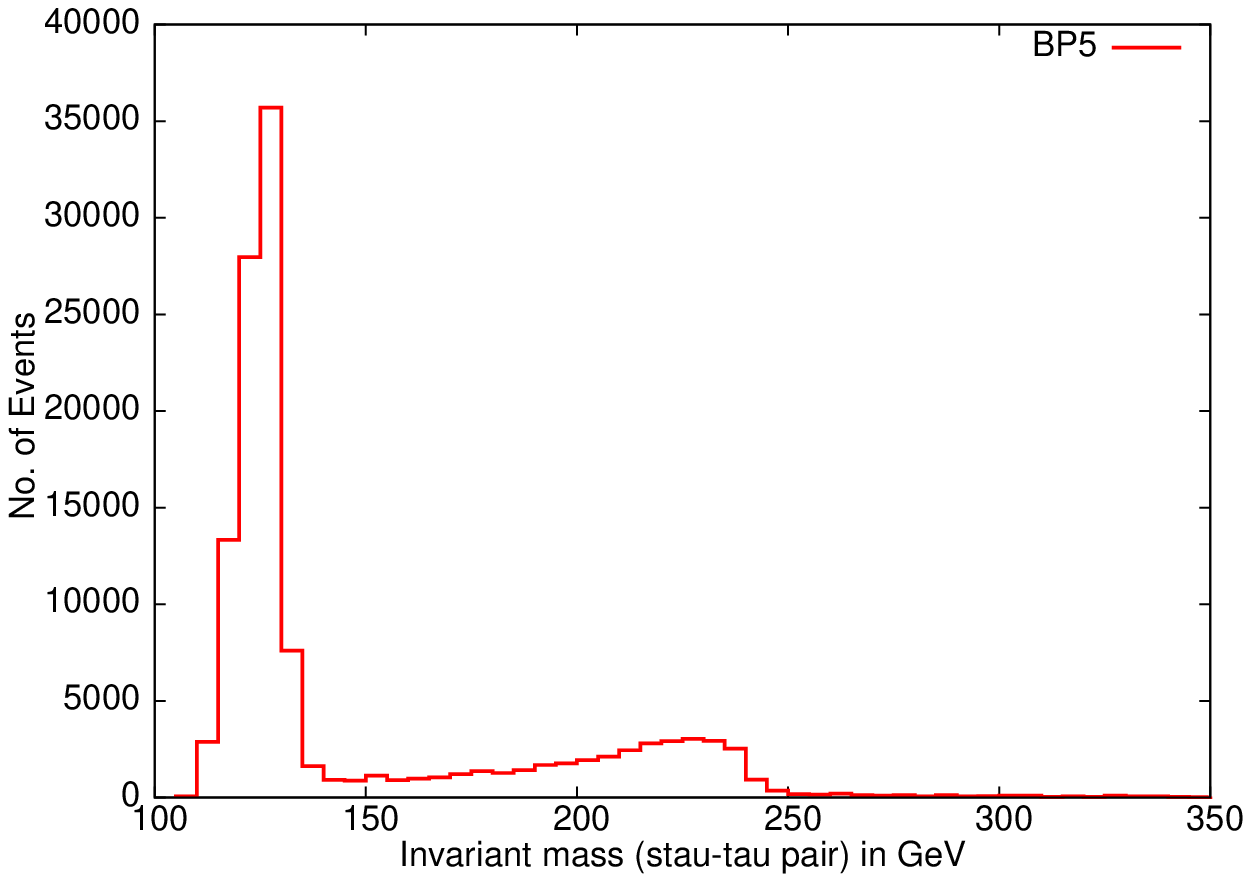,width=7.0cm,height=6.0cm,angle=-0}
\hskip 20pt \epsfig{file=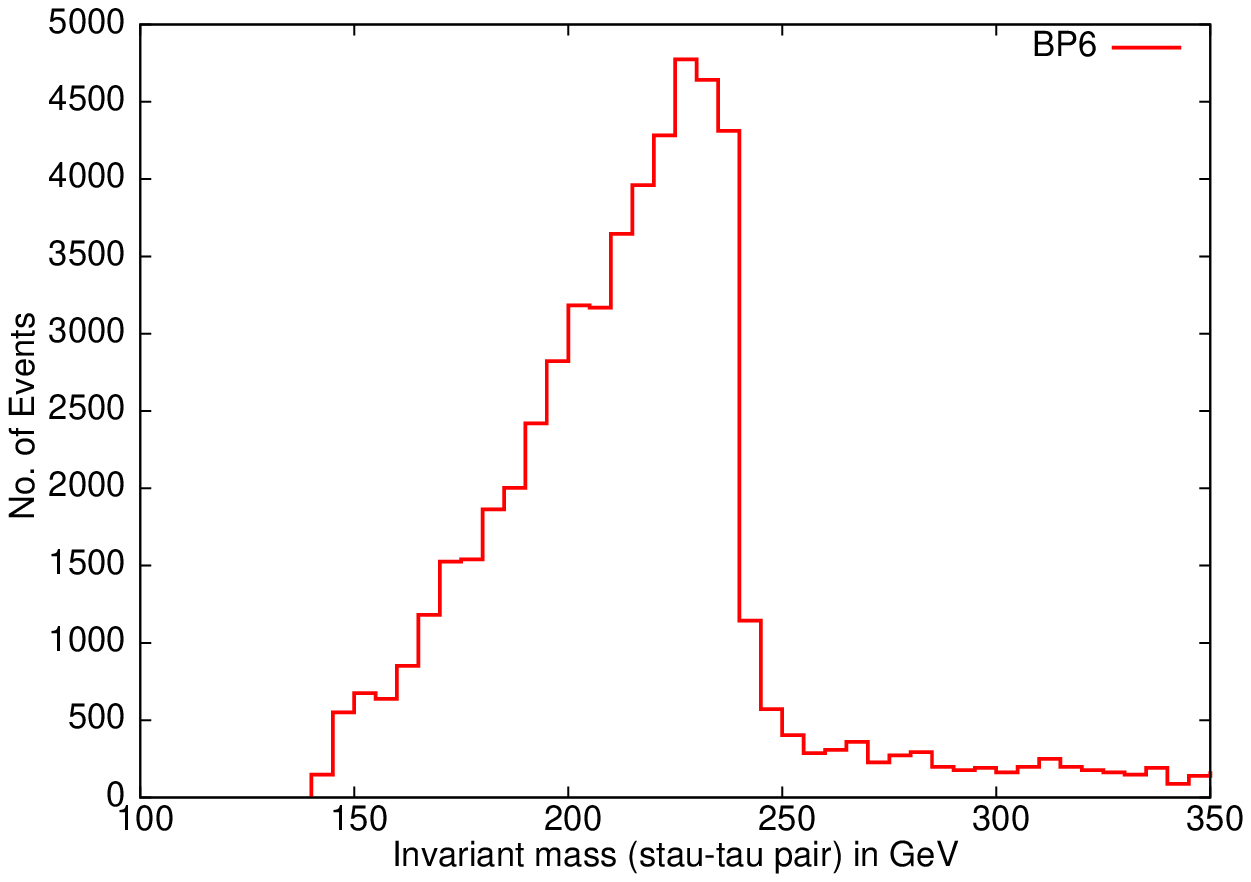,width=7.0cm,height=6.0cm,angle=-0}}
\vskip 20pt
\caption{\small \it {$M_{\stau\tau_j}$ distribution for all the benchmark points. BP1, BP2 and
BP3 show only the $\chi^0_1$ peak. Both the $\chi^0_1$ and $\chi^0_2$ peaks are visible
for BP4 and BP5, while BP6 displays only the $\chi^0_2$ peak.}} 
\end{center}
\end{figure}

Next, for each event that we record, we look at a $\stau$ and a
$\tau$-jet of opposite signs. The invariant mass distribution of this
$\stau\tau_j$ pair displays a peak whose location, although not
precisely telling us about the parent neutralino, is still in the
vicinity of the mass values. Thus, by observing these distributions
(Figure 2) one often is able to tell whether it is a $\chi^0_1$ or a
$\chi^0_2$, once one simultaneously uses information obtained from the
$M_{eff}$ distribution.

As has already been noted in \cite{Biswas:2009zp}, the mass of either 
$\chi^0_1$ or
$\chi^0_2$ or both can be reconstructed in this scenario, depending on
one's location in the parameter space. Once a peak in the $\tau\stau$
invariant mass is located, the next step is to check whether
$|M_{\stau-\tau_j}-m_{\chi^0_i}|<0.1.m_{\chi^0_i}$, where
$m_{\chi^0_i}$ is either one ({\it or the only one}) of the two
lightest neutralinos deemed reconstructible in the corresponding
region.  The mass of that neutralino is used in equation 9. If this
equality is not satisfied for either neutralino or the only one
reconstructed, then the event is not included in the analysis.
 

\subsection{SM backgrounds and cuts}

The final state we have considered, namely, $\tau_j+2\stau {\rm
(opposite-sign~charged~track)}+E_{T}\miss+X$, suffers from several SM
background processes. This is because charged tracks in the muon
chamber due to the presence of quasistable charged particle can be
faked by muons. Such faking is particularly likely for ultra-relativistic
particles, for which neither the time-delay measurement nor the degree of
ionisation in the inner tracking chamber is a reliable discriminator.
The dominant contributions come from
the following subprocesses:

\begin{figure}[htbp]
\centerline{\epsfig{file=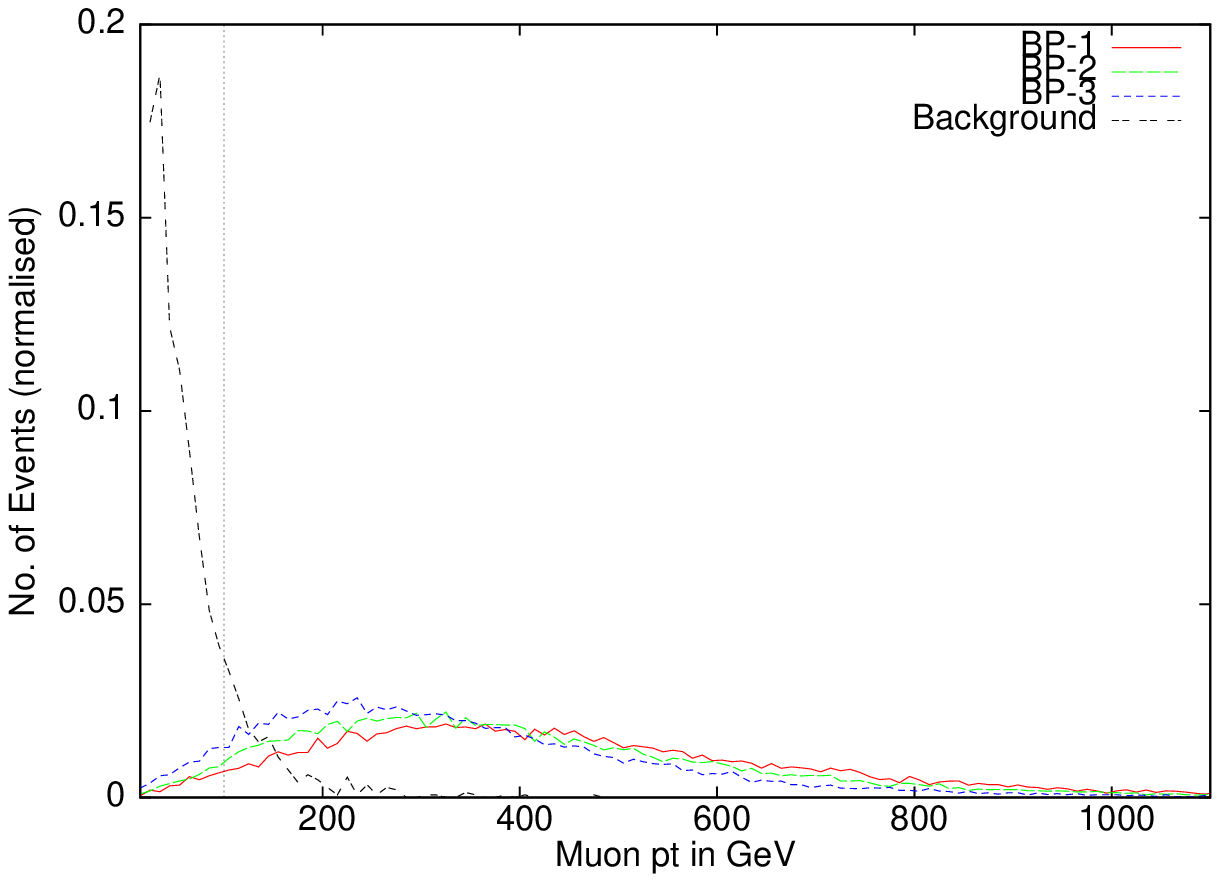,width=8.0cm,height=7.5cm,angle=-0}
\hskip 20pt \epsfig{file=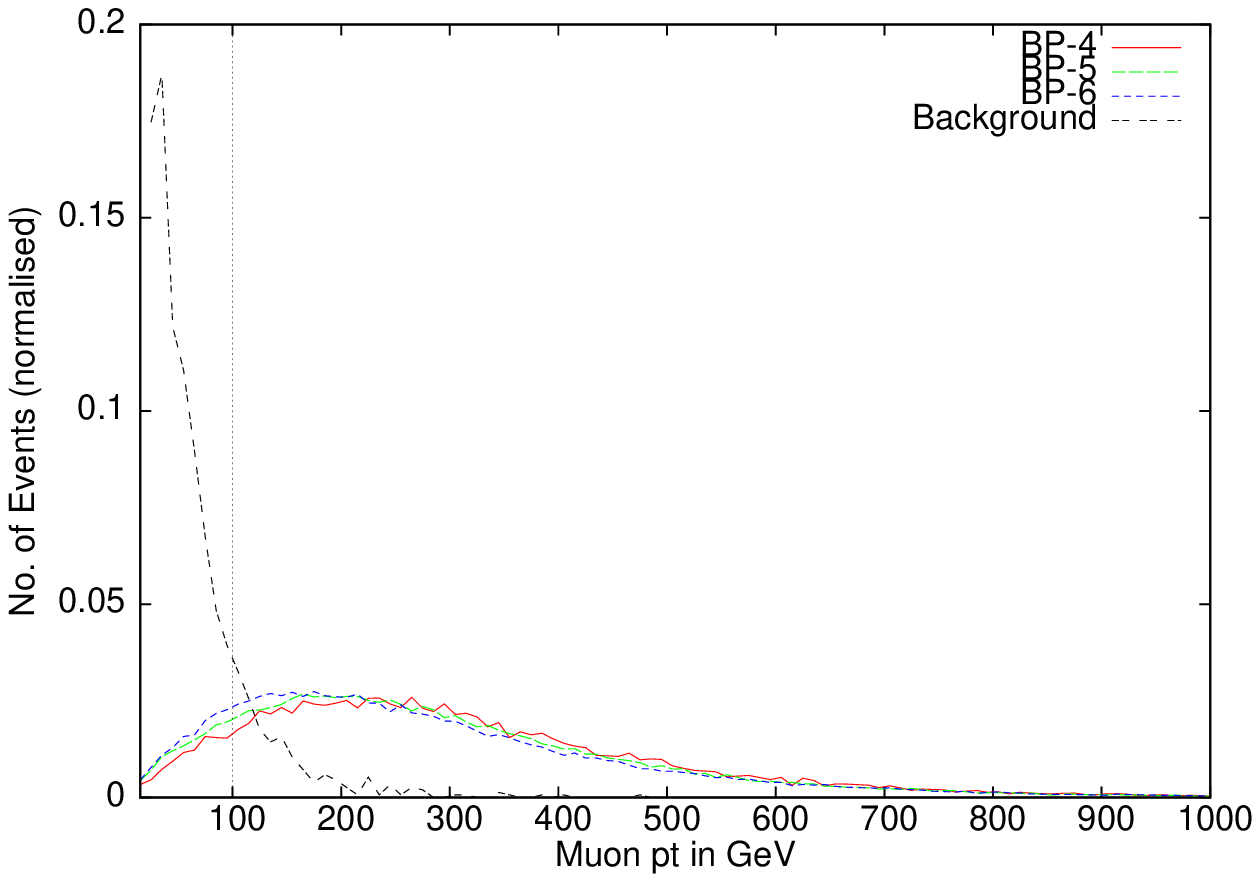,width=8.0cm,height=7.5cm,angle=-0}}
\vskip 15pt
\caption{\small \it {$p_T$ distribution (normalised to unity) of the harder muonlike track 
for the signal and the background, 
for all benchmark points. The vertical lines indicate the effects of a $p_T$-cut
at 100 GeV.}} 
\end{figure}
\vspace{-0.5cm}
\begin{figure}[htbp]
\centerline{\epsfig{file=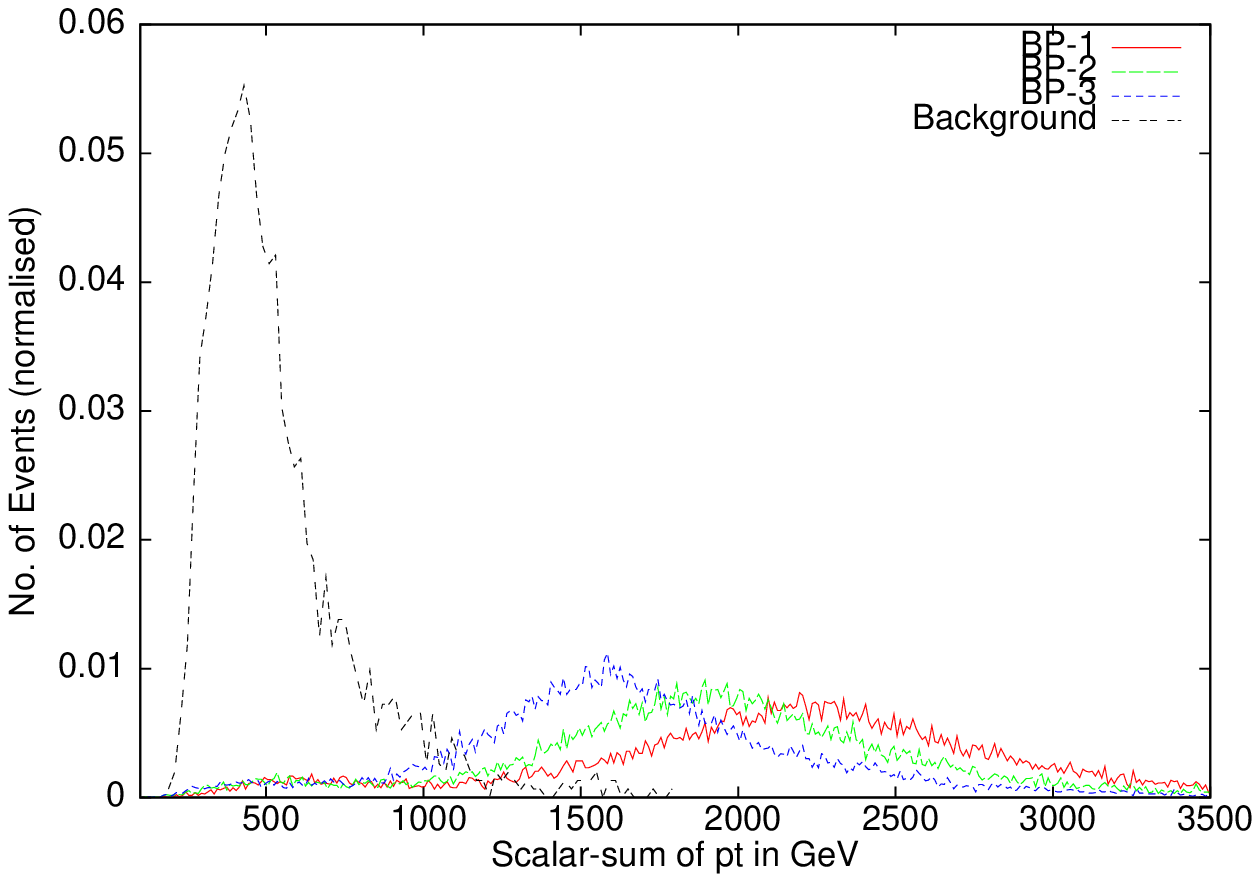,width=8.0cm,height=7.5cm,angle=-0}
\hskip 20pt \epsfig{file=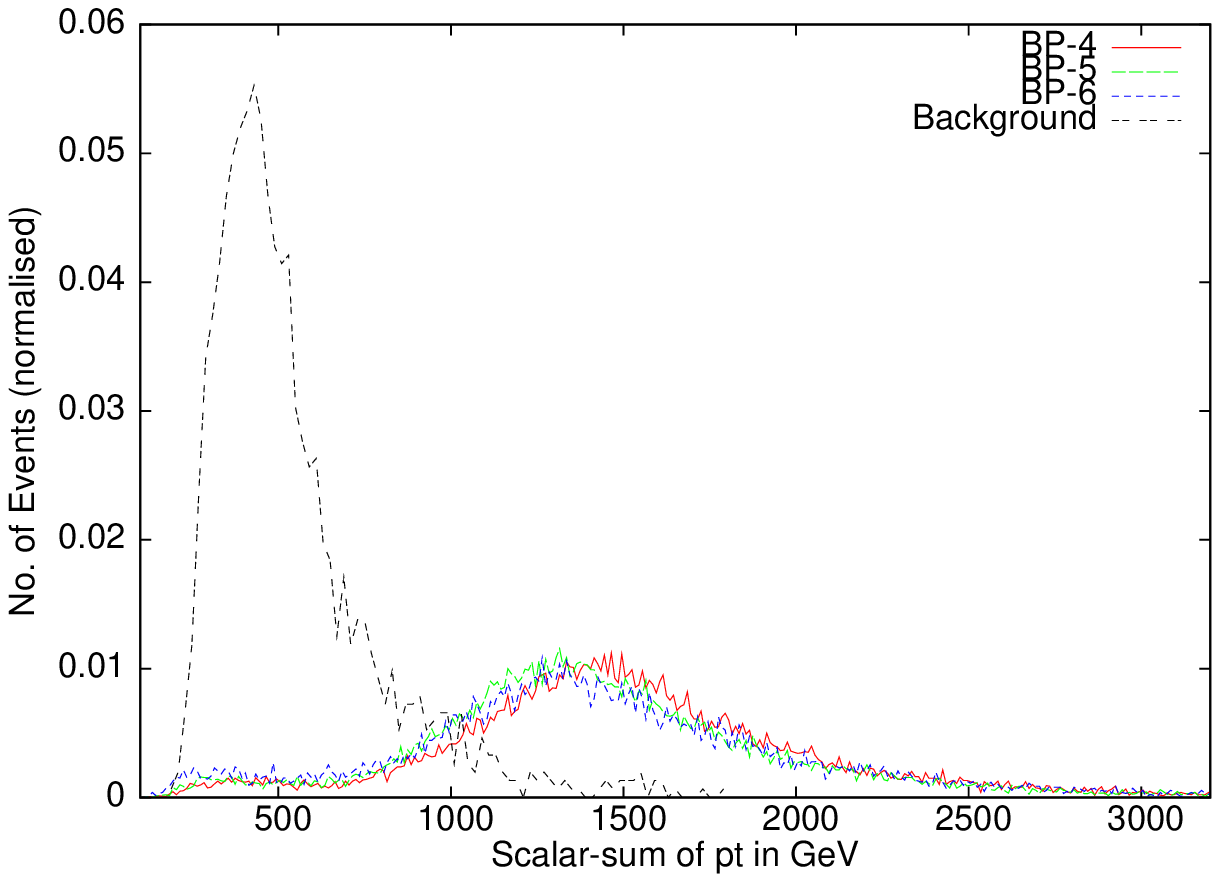,width=8.0cm,height=7.5cm,angle=-0}}
\vskip 15pt
\caption{\small \it {$\Sigma |p_T|$ distribution (normalized to unity) 
for the signal and the background, 
for all benchmark points.}} 
\end{figure}
\vspace{-0.5cm}

\vskip 30pt

\begin{enumerate}
 \item \underline{{\bf $t\bar t$}}: This is a potential
 background for any final state in the context of LHC, due
 to its large production cross-section. In this case $t\bar t\r
 bW^+\bar bW^-$, followed by various combinations of leptonic as well
 as hadronic decays of the $b$ and the $W$, can produce opposite sign
 dimuons and jets.  The jets may emanate from actual taus, but may as
 well be fake. One has an efficiency of 50\% in the former case, and a
 mistagging probability of 1\% in the latter. The $t\bar t$ cross-section 
has been multiplied by a K-factor of 1.8 \cite{ktt}.

 \item \underline{{\bf $b\bar b$}}: This, too, has an
 overwhelmingly large event rate at the LHC. The semileptonic decay of
 both the b's ($b\r c\mu\nu_{\mu}$) can give rise to a dimuon final
 state and any of the associated jets can be faked as tau-jet. Though
 the mistagging probability of a non-tau jet being identified as a
 tau-jet is small, the large cross-section of $b\bar b$ production 
 warrants serious attention to this background.

\item  \underline{{\bf $ZZ$}}: In this case 
any one of the $Z$'s can decays into a dimuon pair ($Z\r \mu\mu$)
while the other one can decays into $\tau\tau$ pair where only one of
the tau can be identified. The hadronic decay of $Z$ and the
subsequent misidentification of any of them as tau-jet is also
possible.

\item   \underline{{\bf $ZW$}}: This SM process also contributes to the 
final state under consideration with
$Z\r \mu\mu$ and $W\r \tau\bar \nu_{\tau}$.

\item \underline{{\bf $ZH$}}: This subprocess can also 
contribute to the final state $\tau_j+2\mu (charged-track)+E_{T}\miss$
where the Higgs decaying into a pair of $\tau$'s, with only one of the
$\tau$ being identified has been considered.
\end{enumerate}

Our chosen event selection criteria have been prompted by all the
above backgrounds.  First of all, we have subjected the
events to the following basic cuts:

\begin{itemize}
\item $ p_{T}^{lep, track} > 10$~GeV
\item $ p_{T}^{hardest-jet} > 75$~GeV 
\item $ p_{T}^{other-jets} > 30$~GeV 
\item $ 40~GeV<\sla{E_T} < 180~ GeV$ 
\item $|\eta| < 2.5$  for Leptons, Jets \& Stau 
\item $\Delta R_{ll}>0.2,~ \Delta R_{lj}>0.4$, where $\Delta R^2=\Delta \eta^2+\Delta \phi^2$
\item $\Delta R_{\stau l}>0.2, ~\Delta R_{\stau j}>0.4$
\item $ \Delta R_{jj} >0.7$
\end{itemize}

\noindent
Though the above cuts largely establish the {\it bona fide} of a
signal event, the background events are too numerous to be effectively
suppressed by them. One therefore has to use the fact that the jets
and stau-tracks are all arising from the decays of substantially
heavy sparticles.  This endows them with added degrees of hardness, as
compound to jets and muons produced in SM process. Thus we can impose
a $p_T$ cut on each track on the muon chamber, and also demand a large
value of the scalar sum of transverse momenta of all the visible final
state particles: 

\begin{itemize}
\item $p_T^{muonlike ~track}> 100~GeV$ 
\item $\Sigma |\vec{p_T}| > 1 ~TeV$ 
\end{itemize}

The justification of these cuts can also be seen from
Figures 3 and 4.
It may be noted that no invariant mass cut on the pair of charged tracks has
been imposed. While such a cut, too, can suppress the dimuon background, we
find it more rewarding to use the scalar sum of $p_T$ cut.

\subsection{SUSY backgrounds}

Apart from the SM backgrounds, SUSY processes in this scenario itself
contribute to the final state $\tau_j+2\stau~+~E_{T}\miss+X$, which
are often more serious than the SM backgrounds. These events will
survive the kinematic cuts listed in the previous subsection, since
they, too, originate in heavy sparticles produced in the
initial hard scattering.  The dominant contributions of this kind come
from:

\begin{enumerate}
 \item \underline{{\bf $\chi^0_i\chi^0_j$} production in cascade decay
 of squarks/gluinos}: This is one of the potentially dangerous
 background where both the $\chi^0_i$'s ($i,j=1,2$) decay into 
 $\stau\tau$-pairs, with only one tau being identified. This then mimics
 our final state in all details with a much higher event rate.

\item \underline{{\bf $\snu_{\tau_L}\chi^0_i$} production 
in cascade decay of squarks/gluinos}: The decay of $\snu_{\tau_L}$ as
part of the cascade produces a $W\stau_1$-pair, while the $\chi^0_i$
decays into a $\stau\tau$-pair to give rise to same final state with
an additional $W$ which then can decay hadronically. 
The $\snu_{\tau L}$ is produced in association with a tau in
large number of events (e.g., $\chi^{\pm}_1 \r \snu_{\tau L}\tau$). 
The $\snu_{\tau_L}$ can also be produced from, say, a  $\chi^0_2$. 
In both of the above situations, a tau-stau pair can be seen
together with another stau track, thus leading to a background event.
\end{enumerate}

The first background can be reduced partially by looking at the
invariant mass distribution of the $\stau$ (having same sign as that
of the {\it identified $\tau$} in the final state) with each jet in
the final state. If this distribution for any particular combinations
falls within $m_{\chi_i}-20<M_{\stau j}<m_{\chi_i}+20$ (where
$m_{\chi_i}=m_{\chi^0_1}~or~\frac {m_{\chi^0_2}}{2}$, depending on
whether $\chi^0_1$ or $\chi^0_2$ is better constructed), 
we have thrown away that event. The reason for
this lies in the observations depicted in Figure 2; the invariant mass
of a $\tau$-induced jet and the $\stau$ which shows a peak close to
the mass of the neutralino from which the $\stau$-$\tau$ pair is
produced.  This has been denoted by Cut
X in Tables 2 and 3. In addition, if the available information on the
effective mass tells us that the $\chi^0_2$ is better reconstructed in
the region, and is produced along with a $\chi^0_1$ with substantial
rate, then a similar invariant mass cut around the $\chi^0_2$ mass
will also be useful. A further cut on the transverse mass distribution
$M^T_{\stau\nu_{\tau}}$ ($> 1.5m_{\chi^0_1}$ or $0.75 m_{\chi^0_2}$) substantially
decreases this background without seriously effecting the signal.

The background of the second kind can in principle be reduced by
vetoing events with additional $W$'s. To identify events with $W$ we
have considered only the hadronic decays of $W$'s. We first observe
the $\Delta R$ separation between the stau (produced in decay of
$\snu_{\tau L}$) and the direction formed out of the vector sum of the
momenta of the two jets produced in $W$ decay. If this separation lies
within $\Delta R=0.8$ and the invariant mass of the two jets lies
within $M_W-20<M_{jj}<M_W+20$ we discard that particular event. In
addition, for a sufficiently boosted $W$, one can have a situation
where the two jets merge to form a single jet. For such a case, we
again look at the stau and each jet within $\Delta R < 0.8$ around
it. The invariant mass of the resultant jet is taken to be 20\% of the
jet energy. The event is rejected if a jet with the mass lies within
$\pm$20 GeV of the $W$-mass.  We have denoted this by Cut Y in Tables
2 and 3. Of course, while it is useful in reducing the
background, a fraction of the signal events also gets discarded in the
process.

\section{Numerical results}

We finally present the numerical results of our study, after imposing
the various cuts for all the benchmark points.  From Tables 2 and 3
one can see that, after demanding a minimum hardness of the charged
track ($p^{track}_T>100$ GeV), together with the cut on the scalar sum of
$p_T$ ($\Sigma|p_T|>1 TeV$), the contribution from the SM processes
get reduced substantially. The cuts X and Y, defined in the previous
subsection, are relatively inconsequential for SM processes. However,
in the process of solving for neutrino momentum in tau decay, most 
of the SM background events gets eliminated on 
demanding the invariant mass of the tau, paired with a oppositely
charged track, to be around the neutralino mass ($m_{\chi^0_1} ~or~
m_{\chi^0_2}$). This is due to the demand that the solution  be
physical, i.e. the fraction $x$ lies between 0 and 1. It is very unlikely to
have admissible solutions for $x$ in SM processes, with the $\tau$-(muon)track
pair invariant mass peaking at $m_{\chi^0_1}$/$m_{\chi^0_2}$. Thus, although
the demand $0<x<1$ is not meant specifically for background
elimination, it is nonetheless helpful in reducing backgrounds. We
have verified that the SM contributions within a bin of
$\pm20$GeV around the reconstructed peak is very small.

As has been already mentioned,
SUSY backgrounds within the model itself is hard to get rid of completely. 
The peak in the transverse mass distribution of the $\stau$-$\nu_{\tau}$
pair get smeared due to such background events (see Figure 4). 
We have already mentioned two suggested cuts,  namely, X and Y, 
which partially reduce these backgrounds. Of these, cut Y suppresses
(by about 15\%) some of the $\snu_{\tau_L}$-$\chi^0_{1/2}$ events, 
as can be seen from  Tables 2 and 3. the
effects of this cut on the other SUSY backgrounds as well as the signal 
are very similar.

Cut X, meant to eliminate mainly the $\chi^0_i$-$\chi^0_j$ background.
Our analysis shows that this cut is rather effective in in this
respect; the event rate is reduced by almost 50\% Surprisingly, it
also reduces the $\snu_{\tau_L}$-$\chi^0_{1/2}$ background by a
considerable amount. The reason for this is the following:
$\snu_{\tau_L}$-$\chi^0_{1/2}$ is produced in cascade decays of
squarks and gluinos and the $\snu_{\tau_L}$ is often produced from a
$\chi^{\pm}_1$ (the branching fraction being 30\% or more in some
BP's). In that case the decay process is $\chi^{\pm}_1 \r
\tau^{\pm}\snu_{\tau_L}$. The tau out of such a $\chi^{\pm}_1$ is
sometimes identified, whereas the tau out of a $\chi^0_{1/2}$
($\chi^0_{1/2} \r \tau^{\pm}\stau^{\mp}$) from the other decay chain
goes untagged. The invariant mass distribution of a track and the jet
coming from an unidentified $\tau$ is clustered around
$m_{\chi^{0}_i}$ ($i$ = 1,2). Thus cut
X turns out to be effective in eliminating this type of background.

After all this effort, however, one still left with background events
which smear the peak in the transverse mass distribution of the
$\stau\nu_{\tau}$-pair. We have to impose an aadditional cut on the
transverse mass distribution to separate the peak from the background
event. This is in the form of the demand
$M^T_{\stau\nu_{\tau}}>\frac{3}{4}m_{\chi^0_2}$, whereby it is
possible to reduce these backgrounds further, as can be seen from
Tables 2 and 3. It is then possible to determine the chargino mass
($m_{\chi^{\pm}_1}$) by looking at the peak, followed by a sharp
descent, in the transverse mass distribution for several benchmark
points.

The transverse mass distributions for different benchmark points are 
shown in Figures 5 and 6. The tau-identification efficiency is
assumed to be 50\% in Figure 5; Figure 6 reflects the improvement
achieved in a relatively optimistic situation when this
efficiency is 70\%. 

\begin{table}[htb]
\footnotesize 
\begin{tabular}{||l||l|l|r|r|r|r||}
\hline
\hline
 {~~BP1} & {~~Signal} & {~~SM} & \multicolumn{4}{c||}{SUSY backgrounds} \\          
 {}  & ($\chi^0_{1/2}-\chi^{\pm}_1$) & {backgrounds} & $\chi^0_1-\chi^0_1$ & $\chi^0_1-\chi^0_2$ 
              & $\chi^0_2-\chi^0_2$ & $\chi^0_{1/2}-\snu_{\tau L}$ \\
\hline 
Basic cuts & 121 & 65588 & 2557 & 62 & 1 & 786 \\
\hline
With $p_T$+$\Sigma{|p_T|}$ Cut & 92 & 202 & 2236 & 49 & 1 & 551 \\
\hline
Cut Y & 83 & 202 & 1969 & 42 & 1 & 433 \\
\hline
Cut X & 58 & 202 & 1130 & 26 & 1 & 244 \\
\hline
$M^T_{\stau\nu_{\tau}}>\frac{3}{4}m_{\chi^0_2}$ & 28 & 10 & 83 & 2 & 0 & 25 \\
\hline
  $|M_{peak}-M^T_{\stau\nu_{\tau}}|\le 20$& 9 & 3 & 7 & 0 & 0  & 2 \\
\hline 
\hline
 {~~BP2} & {~~Signal} & {~~SM} & \multicolumn{4}{c||}{SUSY backgrounds} \\          
 {}  & ($\chi^0_{1/2}-\chi^{\pm}_1$) & {backgrounds} & $\chi^0_1-\chi^0_1$ & $\chi^0_1-\chi^0_2$ 
              & $\chi^0_2-\chi^0_2$ & $\chi^0_{1/2}-\snu_{\tau L}$ \\
\hline 
Basic cuts & 677 & 65588 & 6600 & 390 & 9  & 2157 \\
\hline
With $p_T$+$\Sigma{|p_T|}$ Cut & 492 & 202 & 5552 & 301 & 7 & 1418 \\
\hline
Cut Y & 444 & 202 & 4885 & 262 & 6 & 1106 \\
\hline
Cut X & 336 & 202 & 2675 & 170 & 5 & 605 \\
\hline
$M^T_{\stau\nu_{\tau}}>\frac{3}{4}m_{\chi^0_2}$ & 173 & 3 & 278 & 26 & 1 & 76 \\
\hline
$|M_{peak}-M^T_{\stau\nu_{\tau}}|\le 20$ & 62 & 0 & 33 & 5 & 0 & 11 \\
\hline 
\hline
 {~~BP3} & {~~Signal} & {~~SM} & \multicolumn{4}{c||}{SUSY backgrounds} \\          
 {}  & ($\chi^0_{1/2}-\chi^{\pm}_1$) & {backgrounds} & $\chi^0_1-\chi^0_1$ & $\chi^0_1-\chi^0_2$ 
              & $\chi^0_2-\chi^0_2$ & $\chi^0_{1/2}-\snu_{\tau L}$ \\
\hline 
Basic cuts & 5519 & 65588 & 19400 & 3361 & 170 & 6959 \\
\hline
With $p_T$+$\Sigma{|p_T|}$ Cut & 3571 & 202 & 15181 & 2240 & 98 & 4186 \\
\hline
Cut Y & 3131 & 202 & 13091 & 1924 & 91 & 3231 \\
\hline
Cut X & 2372 & 202 & 6974 & 1192 & 71 & 1679 \\
\hline
$M^T_{\stau\nu_{\tau}}>\frac{3}{4}m_{\chi^0_2}$ & 1189 & 0 & 985 & 205 & 14 & 208 \\
\hline
 $|M_{peak}-M^T_{\stau\nu_{\tau}}|\le 20$ & 523 & 0 & 154 & 46 & 1 & 27 \\
\hline 
\hline
\end{tabular}
\caption{\small \it {Number of signal and background events for
    the $\tau_j+2\stau$ (charged-track)+$E_{T}\miss+X$ final state, 
      considering all SUSY processes, for BP1, BP2 and BP3 at an 
integrated luminosity 300 $fb^{-1}$ assuming tau identification 
efficiency $\epsilon_{\tau}=50\%$.}}
\label{tab:2}
\end{table}

\begin{table}[htb]
\footnotesize 
\begin{tabular}{||l||l|l|r|r|r|r||}
\hline
\hline
 {~~BP4} & {~~Signal} & {~~SM} & \multicolumn{4}{c||}{SUSY backgrounds} \\          
 {}  & ($\chi^0_{1/2}-\chi^{\pm}_1$) & {backgrounds} & $\chi^0_1-\chi^0_1$ & $\chi^0_1-\chi^0_2$ 
              & $\chi^0_2-\chi^0_2$ & $\chi^0_{1/2}-\snu_{\tau L}$ \\
\hline 
Basic cuts & 18194 & 65588 & 33076 & 10618 & 886 & 10613 \\
\hline
With $p_T$+$\Sigma{|p_T|}$ Cut & 10697 & 202 & 24475 & 6342 & 439 & 5713 \\
\hline
Cut Y & 9431 & 202 & 21100 & 5431 & 368 & 4436 \\
\hline
Cut X & 4875 & 202 & 7583 & 2132 & 157 & 1480 \\
\hline
$M^T_{\stau\nu_{\tau}}>\frac{3}{4}m_{\chi^0_2}$ & 2345 & 0 & 1274 & 439 & 41 & 231 \\
\hline
  $|M_{peak}-M^T_{\stau\nu_{\tau}}|\le 20$ & 1076 & 0 & 254 & 114 & 13  & 58 \\
\hline 
\hline
 {~~BP5} & {~~Signal} & {~~SM} & \multicolumn{4}{c||}{SUSY backgrounds} \\          
 {}  & ($\chi^0_{1/2}-\chi^{\pm}_1$) & {backgrounds} & $\chi^0_1-\chi^0_1$ & $\chi^0_1-\chi^0_2$ 
              & $\chi^0_2-\chi^0_2$ & $\chi^0_{1/2}-\snu_{\tau L}$ \\
\hline 
Basic cuts & 34489 & 65588 & 39521 & 19574 & 1976  & 12039 \\
\hline
With $p_T$+$\Sigma{|p_T|}$ Cut & 18748 & 202 & 28329 & 10827 & 958 & 5953 \\
\hline
Cut Y & 16419 & 202 & 24348 & 9326 & 815 & 4586 \\
\hline
Cut X & 8508 & 186 & 8869 & 3764 & 376 & 1626 \\
\hline
$M^T_{\stau\nu_{\tau}}>\frac{3}{4}m_{\chi^0_2}$ & 4099 & 0 & 1574 & 866 & 144 & 258 \\
\hline
  $|M_{peak}-M^T_{\stau\nu_{\tau}}|\le 20$& 2145 & 0 & 339 & 221 & 52 & 37 \\
\hline 
\hline
 {~~BP6} & {~~Signal} & {~~SM} & \multicolumn{4}{c||}{SUSY backgrounds} \\          
 {}  & ($\chi^0_{1/2}-\chi^{\pm}_1$) & {backgrounds} & $\chi^0_1-\chi^0_1$ & $\chi^0_1-\chi^0_2$ 
              & $\chi^0_2-\chi^0_2$ & $\chi^0_{1/2}-\snu_{\tau L}$ \\
\hline 
Basic cuts & 17146 & 65588 & 14519 & 20756 & 1970 & 4644 \\
\hline
With $p_T$+$\Sigma{|p_T|}$ Cut & 8379 & 202 & 9593 & 11405 & 968 & 2524 \\
\hline
Cut Y & 7326 & 202 & 8004 & 9776 & 778 & 2025 \\
\hline
Cut X & 4204 & 186 & 3837 & 5697 & 374 & 1038 \\
\hline
$M^T_{\stau\nu_{\tau}}>\frac{3}{4}m_{\chi^0_2}$ & 1475 & 0 & 231 & 1783 & 128 & 15 \\
\hline
 $|M_{peak}-M^T_{\stau\nu_{\tau}}|\le 20$ & 774 & 0 & 62 & 569 & 44 & 0 \\
\hline 
\hline
\end{tabular}
\caption{\small \it {Same as in Table-2, but for BP4, BP5 and BP6.}}
\label{tab:2}
\end{table}

\begin{figure}[htbp]
\centerline{\epsfig{file=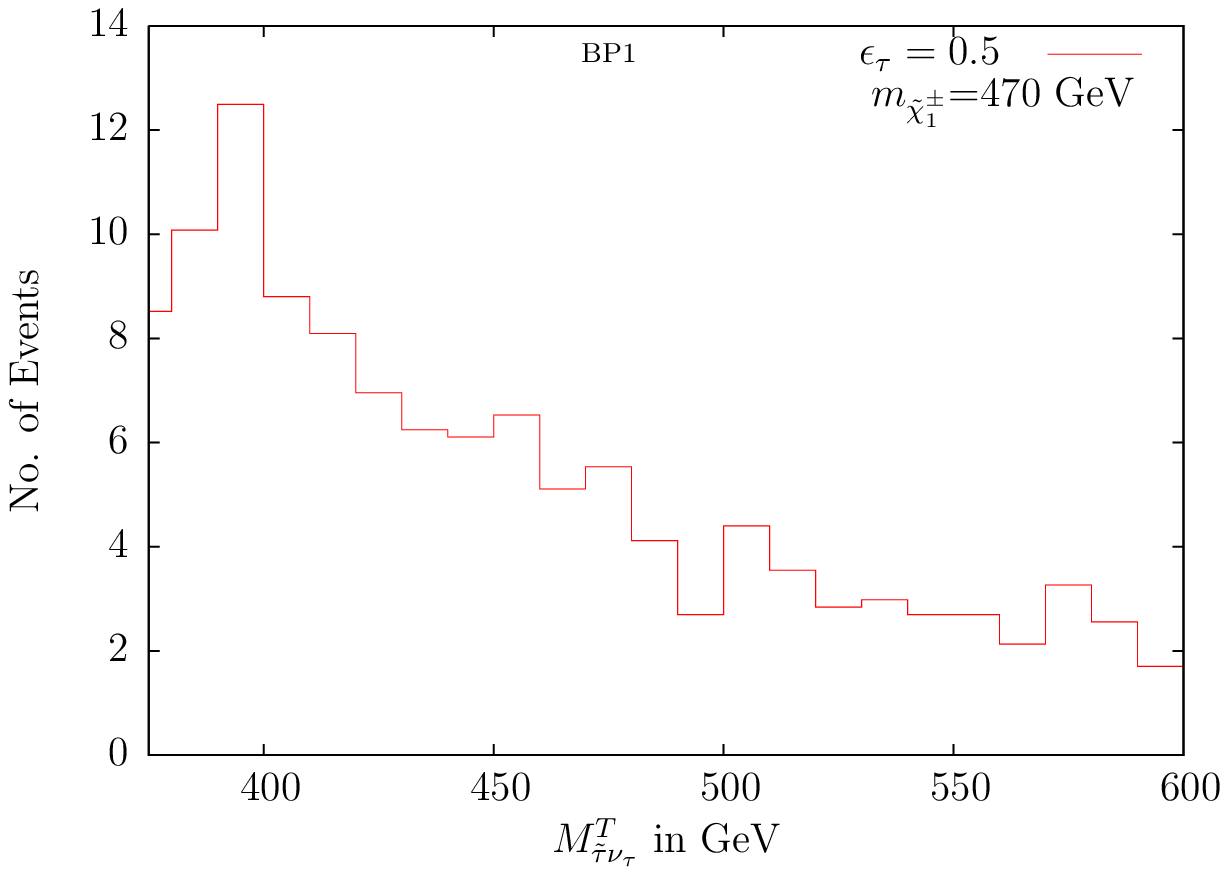,width=7.0cm,height=6.0cm,angle=-0}
\hskip 20pt \epsfig{file=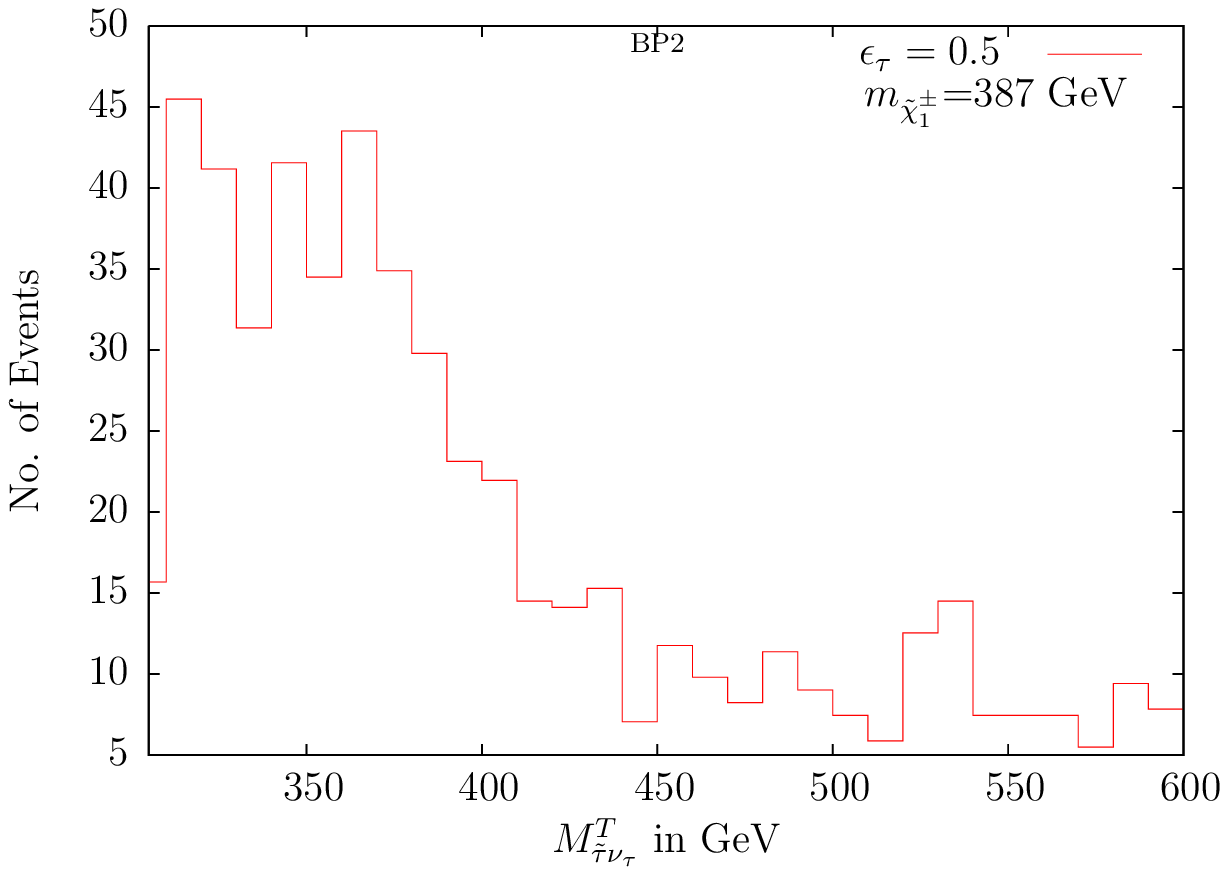,width=7.0cm,height=6.0cm,angle=-0}}
\vskip 10pt
\centerline{\epsfig{file=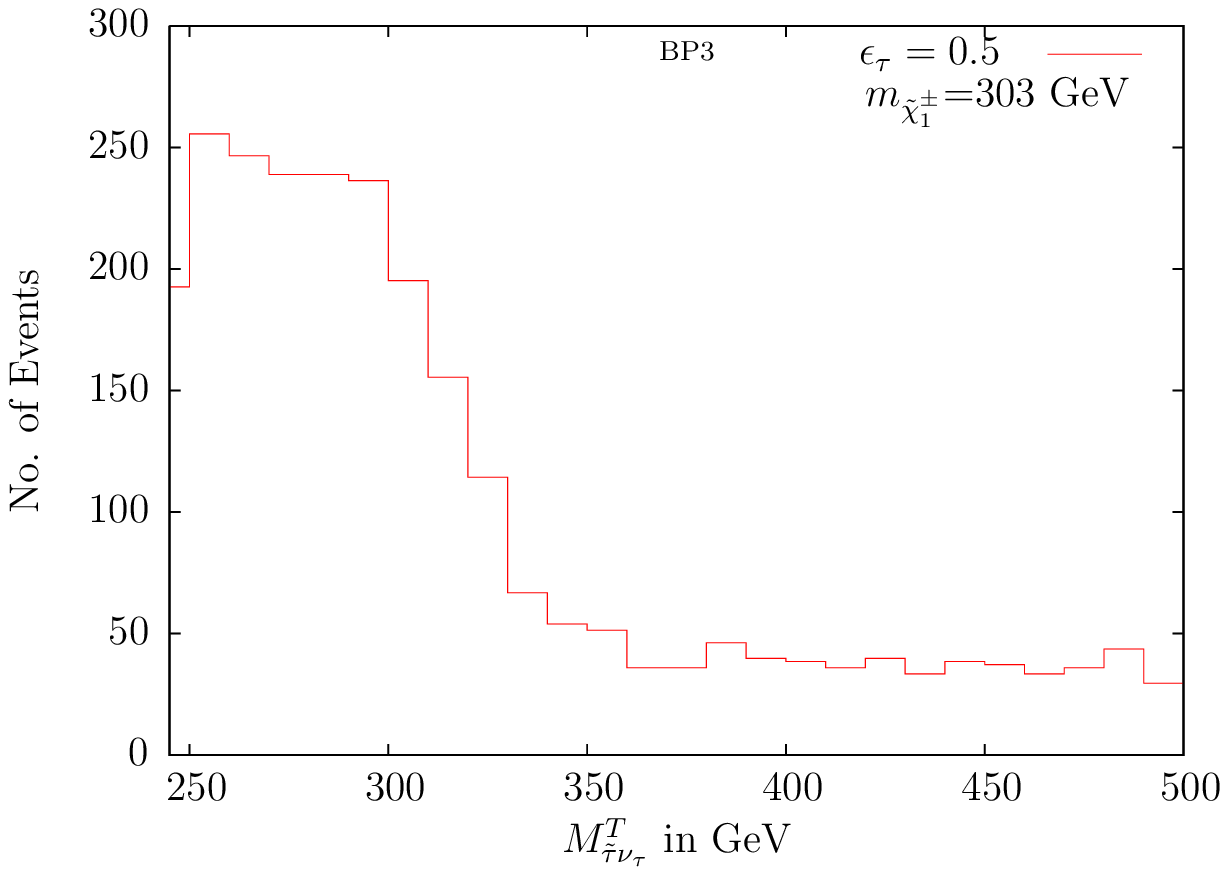,width=7.0cm,height=6.0cm,angle=-0}
\hskip 20pt \epsfig{file=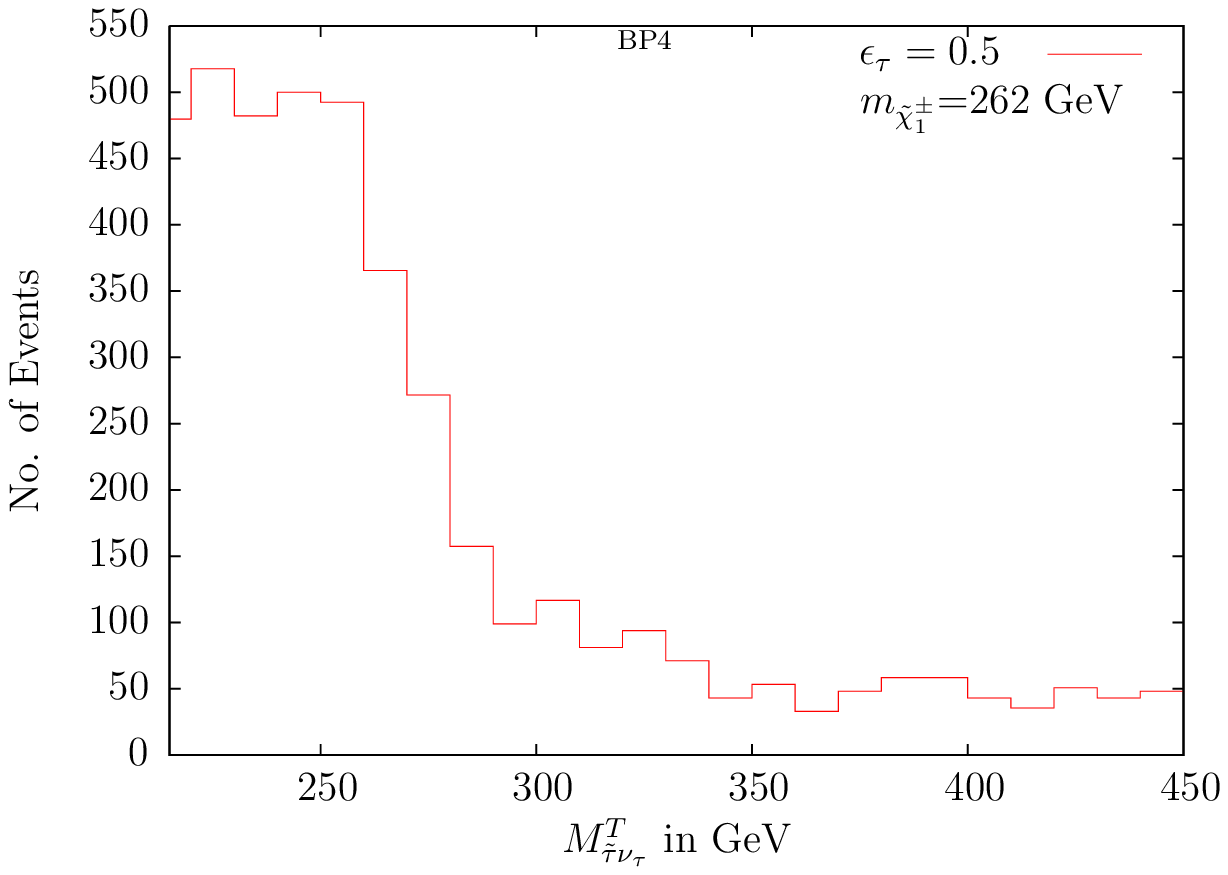,width=7.0cm,height=6.0cm,angle=-0}}
\vskip 10pt
\centerline{\epsfig{file=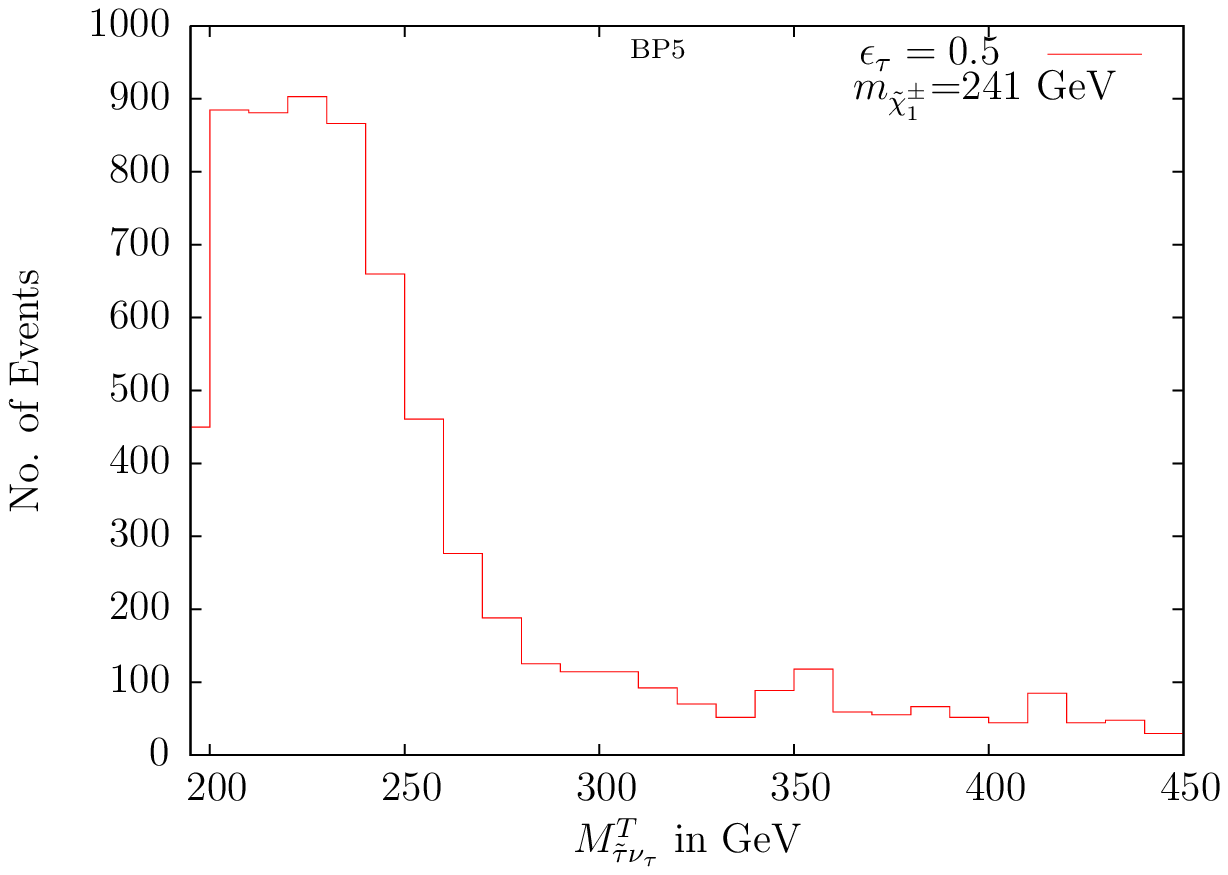,width=7.0cm,height=6.0cm,angle=-0}
\hskip 20pt \epsfig{file=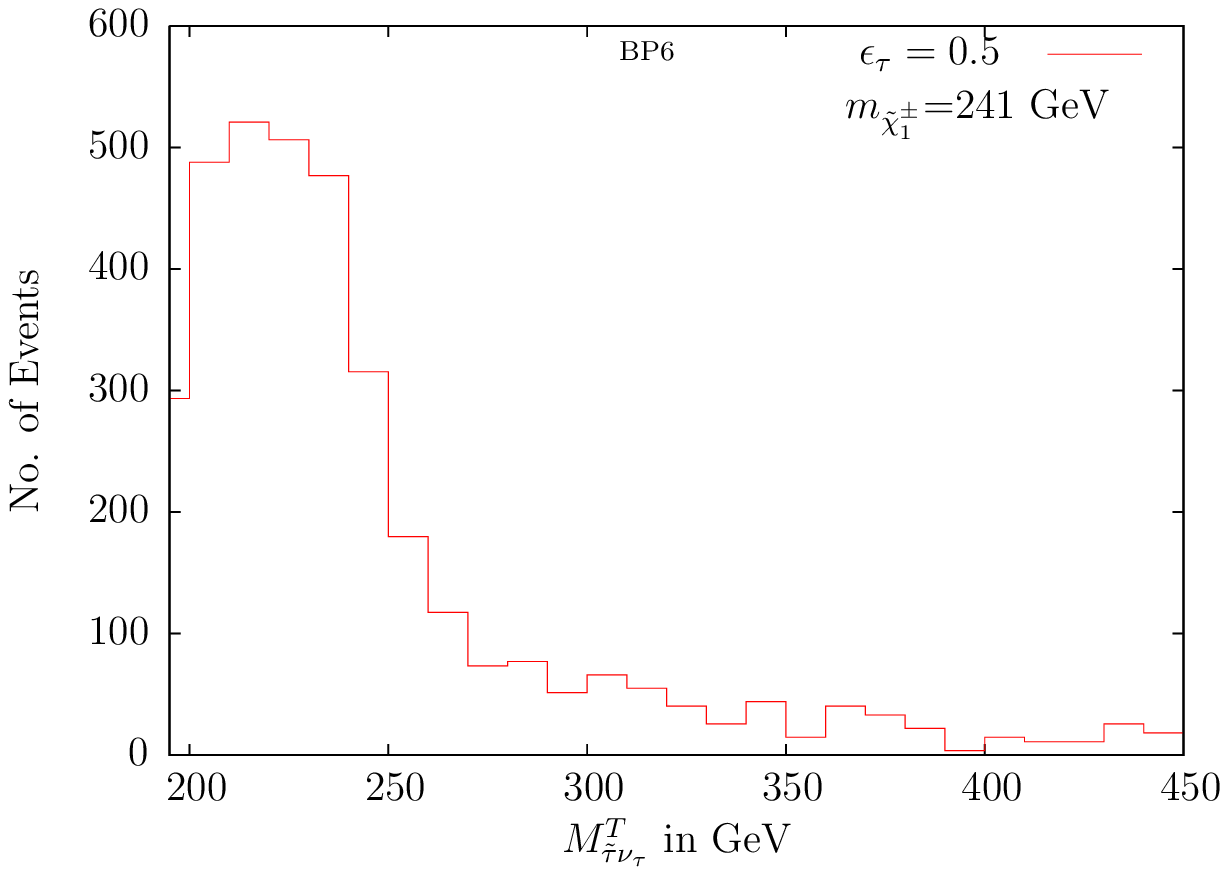,width=7.0cm,height=6.0cm,angle=-0}}
\vskip 20pt
\caption{\small \it {The transverse mass $\stau\nu_{\tau}$-pair from 
chargino deacy described in the text, for all the benchmark points with tau 
identification efficiency $(\epsilon_{\tau})=50\%$.}} 
\end{figure}

\begin{figure}[htbp]
\centerline{\epsfig{file=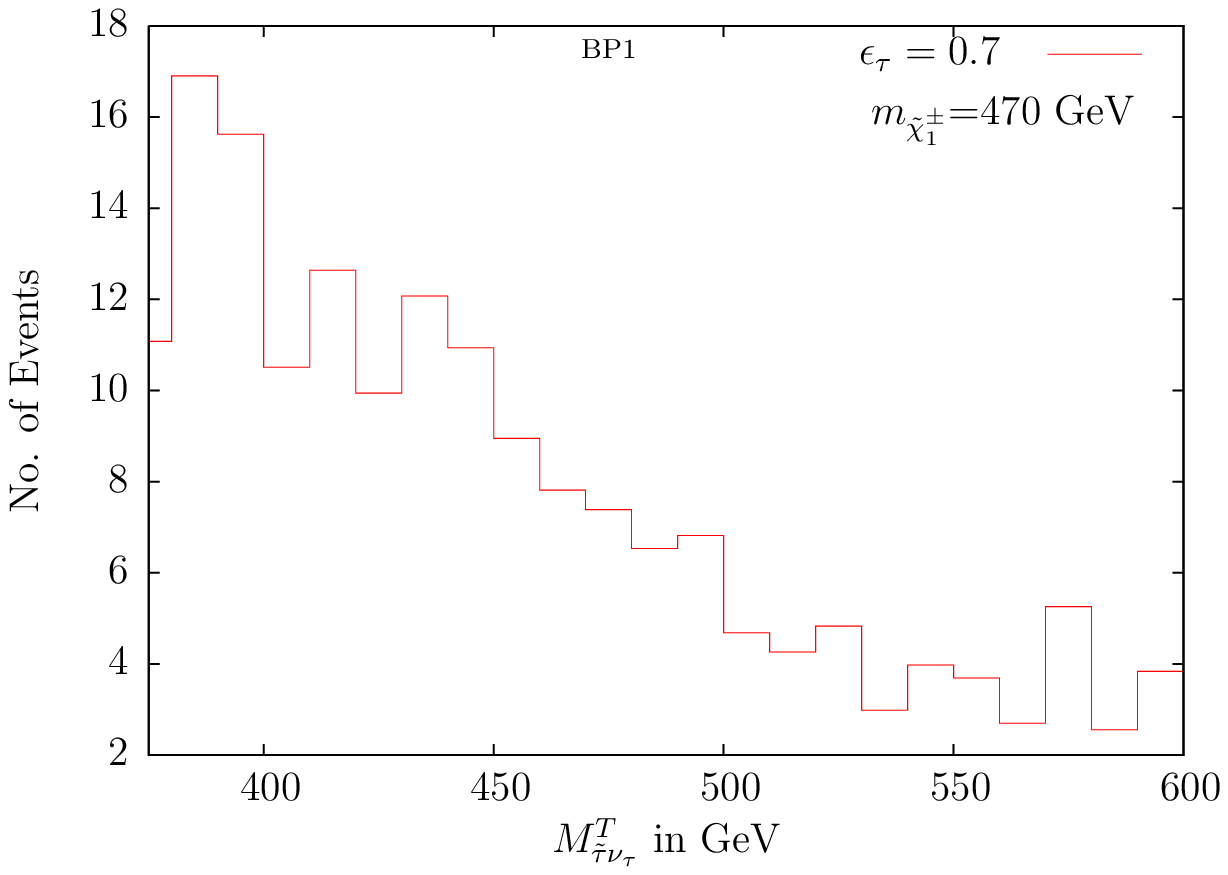,width=7.0cm,height=6.0cm,angle=-0}
\hskip 20pt \epsfig{file=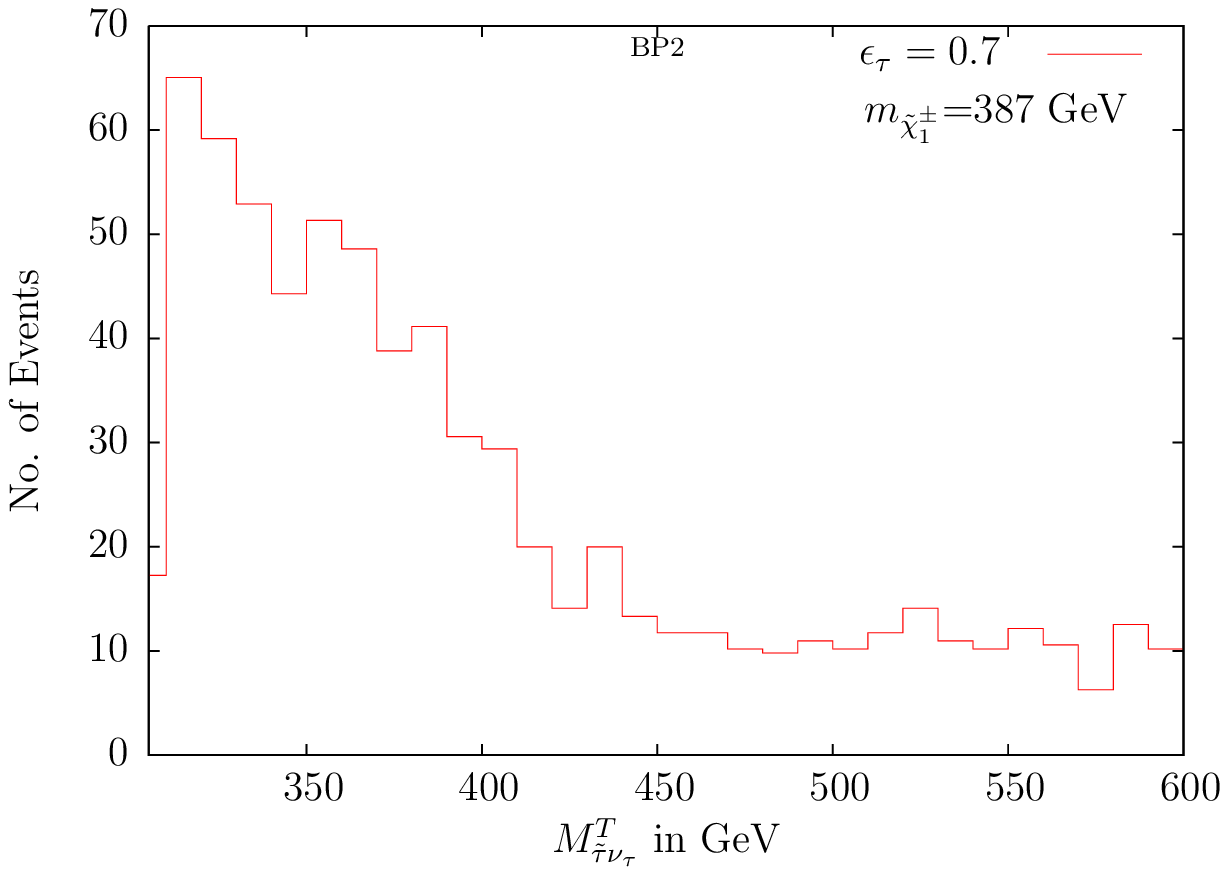,width=7.0cm,height=6.0cm,angle=-0}}
\vskip 10pt
\centerline{\epsfig{file=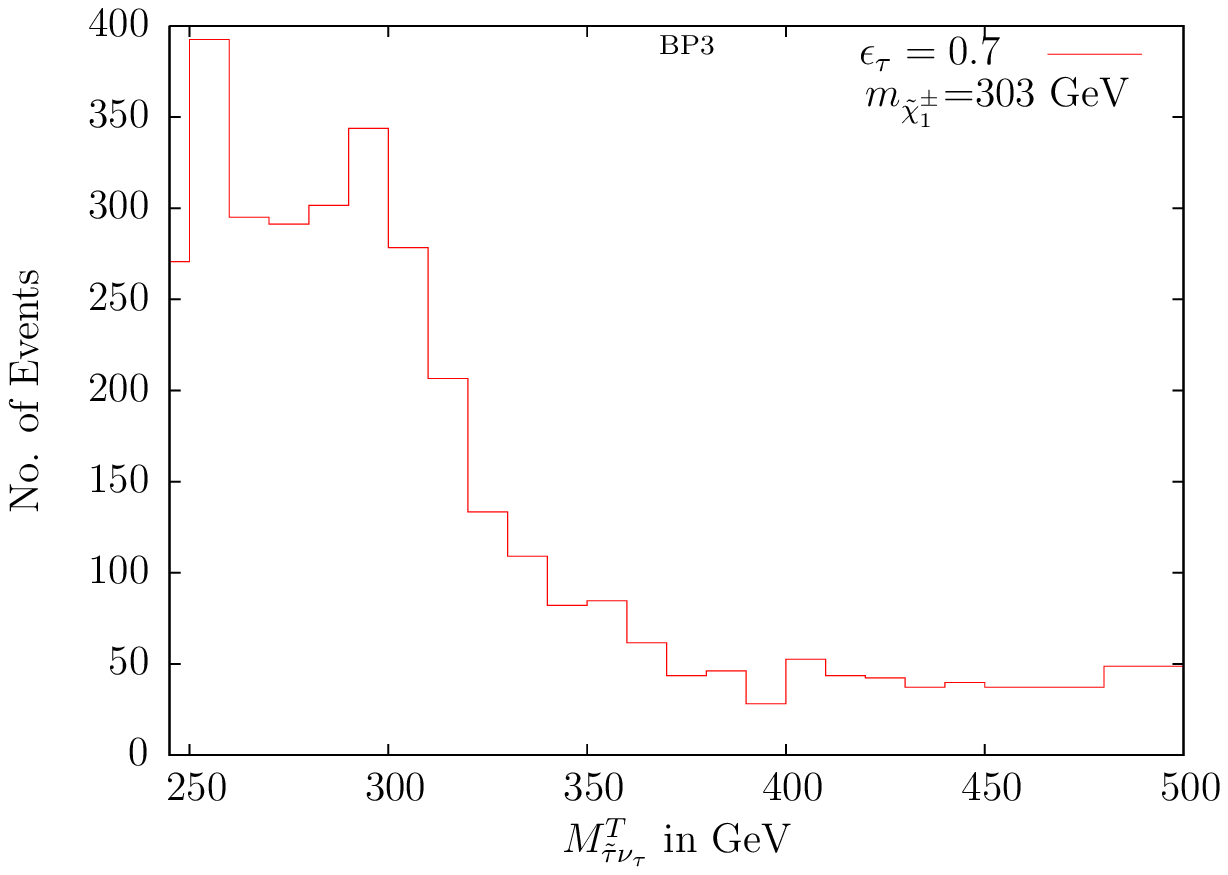,width=7.0cm,height=6.0cm,angle=-0}
\hskip 20pt \epsfig{file=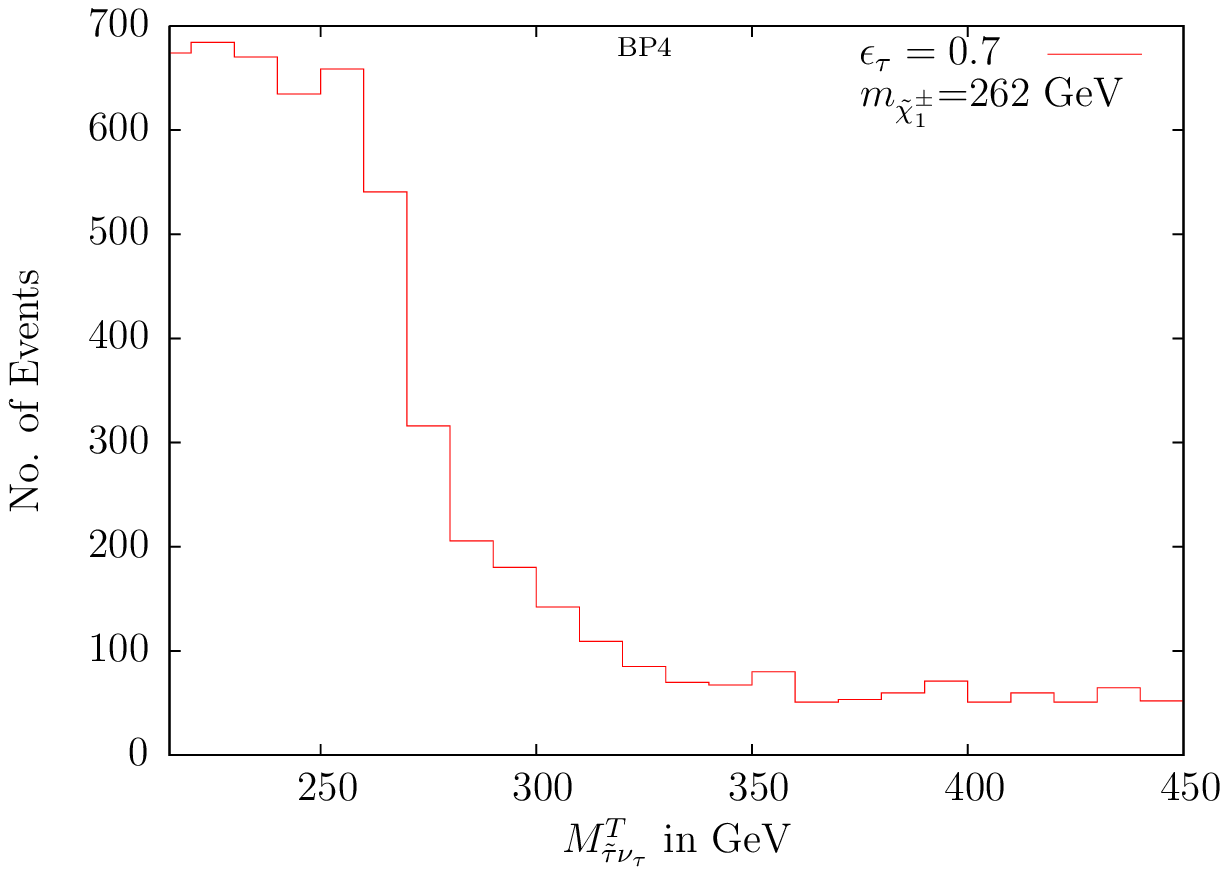,width=7.0cm,height=6.0cm,angle=-0}}
\vskip 10pt
\centerline{\epsfig{file=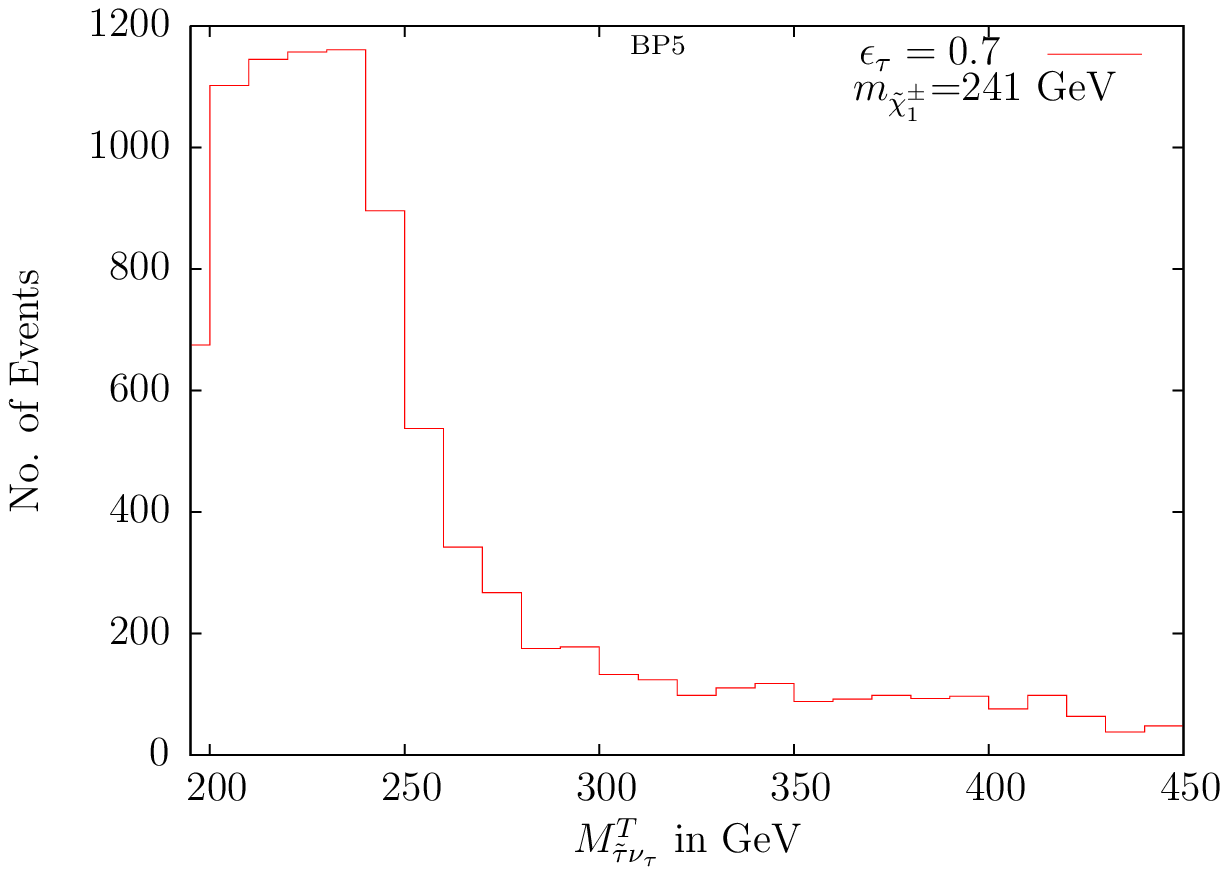,width=7.0cm,height=6.0cm,angle=-0}
\hskip 20pt \epsfig{file=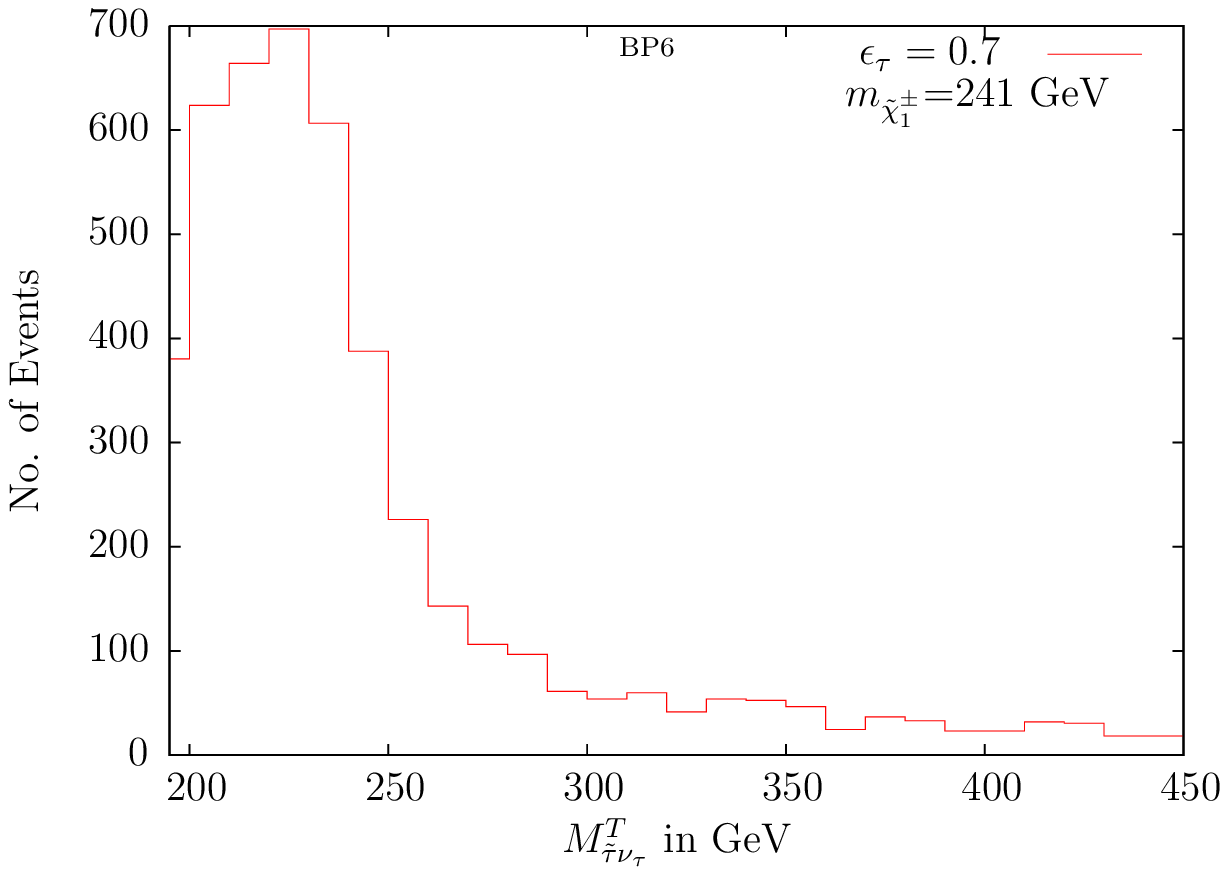,width=7.0cm,height=6.0cm,angle=-0}}
\vskip 20pt
\caption{\small \it {Same as in Figure 5, but with tau identification 
efficiency $(\epsilon_{\tau})=70\%$.}} 
\end{figure}
\vspace{1.5cm}

At BP1 the statistics is very poor and we have realtively 
few events within a bin of 40 GeV around $m_{\chi^{\pm}_1}$.
One has about 50\% of the events
coming from other SUSY processes (Table 2). Also the peak is not clearly
visible due to the presence
of a large number of $\chi^0_{1} \chi^0_{1}$ events, even after imposing
the $M^T_{\stau\nu_{\tau}}$ $>\frac{3}{4}m_{\chi^0_2}$ cut. 

The situation is similar for BP2 as
well. In BP3 and BP4 the peak followed by a sharp fall is considerably
more distinct, from which one can extract the value of
$m_{\chi^{\pm}_1}$. 

For BP5 and BP6 the contamination due to the SUSY background is found
to be small compared to the other benchmark points.

The $\chi^{\pm}_1$ production rate in cascade decays of squarks/gluinos
is also higher there. Hence the transverse mass distribution shows a distinct
peak, from which a faithful reconstruction of chargino mass is possible.

As a comparison between Figures 5 and 6 shows, the prospect can be
improved noticeably if one has a better tau identification efficiency
($\epsilon_{\tau}=70\%$). In such a case, the background from
$\chi^0_{1}$-$\chi^0_{1}$/$\chi^0_{1}$-$\chi^0_{2}$/$\chi^0_{2}$-$\chi^0_{2}$
is less severe compared to the case where the 
tau identification efficiency is 50\%.

From Figures 5 and 6, one can also see some small peaks in the
$M^T_{\stau\nu_{\tau}}$ distribution with very few event rate, in the
region where $M^T_{\stau\nu_{\tau}}>m_{\chi^0_2}$. This can be
attributed to those events where a $\chi^0_{3}$ or a $\chi^0_{4}$ decays
into a $\stau\tau$ pair, and also to the production of the heavier chargino.

\section{Summary and conclusions}

We have considered a SUSY scenario where the LSP is dominated by a right-sneutrino
state, while a dominantly right-chiral stau is the NLSP. The stau, being
stable on the length scale of collider detectors, gives rise to
charged tracks, the essence of SUSY signal in such a scenario. It is also
shown that such a spectrum follows naturaly from a high-scale scenario
of universal scalar and gaugino masses. 

We have extended our earlier study on the mass reconstruction of
non-strongly interacting superparticles in such cases, by considering
final states resulting from the decays of a $\chi^{\pm}_1 \chi^{0}_{1(2)}$ pair
in SUSY cascades. The final state under consideration is 
 $\tau_j+2\stau {\rm (opposite-sign~charged~tracks)}+E_{T}\miss+X$. We have
systematically developed a procedure for identifying the contribution to
$\vec{p_{T}\miss}$ from the neutrino produced in $\chi^{\pm}_1$-decay, together
with a quasi-stable stau. Once this is possible, the transverse mass distribution
of the corresponding $\stau - \nu_{\tau}$ pair can be extracted from data at the
LHC, and a sharp edge in that distribution yields information on the chargino mass.
While eliminating the SM backgrounds in this process is straightforward, we have
suggested ways of minimising the contamination of the relevant final state
from competing processes in the same SUSY scenario.
Selecting a number of benchmark points in the parameter space,
we show in which regions the above procedure works. In cases where it does not, 
the main causes
of failure are identified as the overwhelmingly large contribution
from $\chi^{0}_1$-pairs, and, for example, in the first two benchmark
points, somewhat poor statistics. The other important issue is the
differentiation between the  $\chi^{0}_1$  and the $\chi^{0}_2$ produced
in association with the $\chi^{\pm}_1$. For this, we make use of the 
assumption of gaugino universality as well as the information extacted from
the effective mass distribution in SUSY processes.

To conclude, the existence of quasi-stable charged particles,
a possibility not too far-fetched,  
opens a new vista in the reconstruction of superparticle masses. 
We have repeatedly suggested utilisation of this
facility in our works on gluino\cite{Choudhury:2008gb} and neutralino\cite{Biswas:2009zp} 
mass reconstruction. The present work underscores a relatively 
arduous task in this respect, in obtaining transverse
mass edges in chargino decays. In spite of rather challenging
obstacles from underlying SUSY processes, we demonstrate the
feasibility of our procedure, which is likely to be enhanced
by improvement in, for example, the W-and tau-identification efficiencies.

{\bf Acknowledgment:}  This work was partially supported by funding 
available from the Department of Atomic Energy, Government of India 
for the Regional Centre for Accelerator-based Particle Physics, 
Harish-Chandra Research Institute. Computational work for this study was
partially carried out at the cluster computing facility of
Harish-Chandra Research Institute ({\tt http:/$\!$/cluster.mri.ernet.in}).


\end{document}